\documentclass[final,12pt]{elsarticle}

\usepackage{amssymb}
\usepackage{amsthm}

\usepackage{siunitx}

\usepackage{multirow}

\usepackage{amsmath}
\usepackage{wrapfig}
\usepackage{lscape}
\usepackage{rotating}
\usepackage{epstopdf}
\usepackage{xcolor}
\usepackage{lipsum}
\usepackage{ulem}
\usepackage[export]{adjustbox}
\usepackage{subfiles}
\usepackage{bm}
\usepackage{lineno}
\usepackage{caption}
\usepackage{subcaption}
\usepackage{cancel}
\usepackage{float}
\usepackage{tikz}
\usepackage{url}

\usepackage{caption}
\usepackage{listings}

\definecolor{codegreen}{rgb}{0,0.6,0}
\definecolor{codegray}{rgb}{0.5,0.5,0.5}
\definecolor{codepurple}{rgb}{0.58,0,0.82}
\definecolor{backcolour}{rgb}{0.97,0.97,0.92}

\lstdefinestyle{mystyle}{
    backgroundcolor=\color{backcolour},   
    commentstyle=\color{codegreen},
    basicstyle = \ttfamily\scriptsize,
    keywordstyle=\color{magenta},
    numberstyle=\tiny\color{codegray},
    stringstyle=\color{codepurple},
    breakatwhitespace=false,         
    breaklines=true,                 
    captionpos=b,                    
    keepspaces=true,                    
    numbersep=3pt,                  
    showspaces=false,                
    showstringspaces=false,
    showtabs=false,                  
    tabsize=2
}

\definecolor{fgwhite}{rgb}{1,1,1}     
\definecolor{fgred}{rgb}{0.8,0,0}     
\definecolor{fgorange}{rgb}{0.93,0.53,0.18}     
\definecolor{fgpurple}{rgb}{0.55,0.1,0.6}     
\definecolor{fggreen}{rgb}{0,0.5,0}     
\definecolor{bggreen}{rgb}{0.8,1,0.8}     
\definecolor{fgblue}{rgb}{0,0,0.7}     
\definecolor{bgblue}{rgb}{0.9,0.9,1}     
\definecolor{fgclay}{rgb}{0.51,0.25,0.04}     

\newcommand{\Tocite}[1]{{{\textcolor{pink}{\bf [CITER]~}}}}

\newcommand{\TBD}[1]{{{\textcolor{magenta}{\textbf{To be done}}}}}

\begin{document}
%
%
%
%
%
\title{\textcolor{black}{Multi-compartment poroelastic models of perfused biological soft tissues: implementation in FEniCSx} }
%
%
\author[1,2,3]{T. Lavigne}
\author[1]{St\'ephane Urcun}
\author[2]{Pierre-Yves Rohan}
\author[3]{Giuseppe Sciumè}
\author[4]{Davide Baroli \corref{cor}}
\ead{davide.baroli@usi.ch}
\author[1,5,6]{Stéphane P.A. Bordas}
%
%
%
\address[1]{Institute of Computational Engineering, Department of Engineering, University of Luxembourg, 6, avenue de la Fonte, Esch-sur-Alzette, L-4364, Luxembourg}
\address[2]{Arts et Metiers Institute of Technology, IBHGC, 151 bd de l’hopital, Paris, 75013, France}
\address[3]{Arts et Metiers Institute of Technology, Univ. of Bordeaux, CNRS, Bordeaux INP, INRAE, I2M Bordeaux, Avenue d'Aquitaine, Pessac, 33607, France}
\address[4]{Università della Svizzera Italiana, Euler Institute, Lugano, Swiss}
\address[5]{Clyde Visiting Fellow, Department of Mechanical Engineering, The University of Utah, Salt Lake City, United States}
\address[6]{Department of Medical Research, China Medical University Hospital, China Medical University, Taichung, Taiwan}

\begin{abstract}
Soft biological tissues demonstrate strong time-dependent and strain-rate mechanical behavior, arising from their intrinsic visco-elasticity and fluid-solid interactions {(especially at sufficiently large time scales)}. The time-dependent mechanical properties of soft tissues influence their physiological functions and {are linked} to several pathological processes. Poro-elastic modeling represents a promising approach because it allows the integration of multiscale/multiphysics data to probe biologically relevant phenomena at a smaller scale and embeds the {relevant} mechanisms at the larger scale. The implementation of {multi-phasic flow poro-elastic models} however is a complex undertaking, requiring extensive knowledge. The open-source software FEniCSx Project provides a novel tool for the automated solution of partial differential equations by the finite element method. This paper aims to provide the required tools to model the mixed formulation of poro-elasticity, from the theory to the implementation, within FEniCSx. Several benchmark cases are studied. A column under confined compression conditions is compared to the Terzaghi analytical solution, using the L2-norm. An  implementation of poro-hyper-elasticity is proposed. A bi-compartment column is compared to previously published results (Cast3m implementation). For all cases, accurate results are obtained in terms of a normalized Root Mean Square Error (RMSE). Furthermore, the FEniCSx computation is found three times faster than the legacy FEniCS one. The benefits of parallel computation are also highlighted.
\end{abstract}



\begin{highlights}
\item Implementation of poro-elastic formulations within FEniCSx.
\item Single and double compartment columns are modeled.
\item Elastic and hyper-elastic solid scaffold are computed.
\item Fast and accurate computation is obtained.
\end{highlights}

\maketitle


\section{Introduction}
\label{sec:introduction}

Numerous biomechanical problems aim to reproduce the behavior of a deformable solid matrix that experiences flow-induced strain such as the brain (\citet{Budday_2019, HosseiniFarid2020, Franceschini2006, Urcun_2022}), muscle tissues (\citet{Lavigne2022b}), tumors (\citet{Scium2013, Scium2021}, \citet{OFTADEH2018249}), articular cartilages (\citet{ATESHIAN20091163}) and lumbar inter-vertebral discs (\citet{Argoubi1996}). The time-dependent mechanical properties of soft tissues influence their physiological functions and {are linked to several pathological processes}. Although a fluid-structure interaction (FSI) problem, the number, and range of fluid flows are generally so vast that the direct approach of a defined boundary between fluid and solid is impossible to apply, as it requires an exponential computational cost at the organ scale with the requirement of extensive data acquisition at the micro-scale. In these cases, homogenization and statistical treatment of the material-fluid system is possibly the only way forward. A prominent technique of this type is that of poro-elasticity.

Extensive studies have shown that poro-elastic models can accurately reproduce the time-dependent behavior of soft tissues under different loading conditions (\citet{Gimnich2019,Argoubi1996,Peyrounette2018,Siddique2017,HosseiniFarid2020,Franceschini2006,Lavigne2022b}). Compared to a visco-(hyper)-elastic formulation (\citet{VanLoocke2009,Simms2012,Wheatley2015, Vaidya2020}), the poro-elastic properties are independent of the sample size (\citet{Urcun_2022}). Furthermore, a poro-elastic approach can integrate multiscale/multiphysics data to probe biologically relevant phenomena at a smaller scale and embed the {relevant} mechanisms at the larger scale (in particular, biochemistry of oxygen and inflammatory signaling pathways), allowing the interpretation of the different time characteristics (\citet{Stphane2020, Scium2013,Scium2021,Gray2014,Mascheroni2016}).

In most commercially available FE software packages used for research in biomechanics (ABAQUS, ANSYS, RADIOSS, etc), pre-programmed material models for soft biological tissues are available. The disadvantage of these pre-programmed models is that they are presented to the user as a "black box". Therefore, many researchers turn to implement their material formulations through user subroutines (the reader is referred, for example, to the tutorial of \citet{FEHERVARY2020103737} on the implementation of a nonlinear hyper-elastic material model using user subroutines in ABAQUS). This task, however, is complex. When documentation is available, these only provide expressions, without any derivations, lack details and background information, making the implementation complex and error-prone. In addition, in case of a custom formulation or the introduction of biochemical equations for example, specific computational skills are required making the task even more challenging. In the end, the use of commercially available FE software packages may limit the straightforward reproducibility of the research by other teams.

The interest in open-source tools has skyrocketed to increase the impact of the studies within the community (for example FEbio, FreeFem, and Utopia \citet{utopia2021, utopiagit}). For Finite Element modeling, the FEniCS project (\citet{FEniCS}) is an Open-Access software that has proven its efficiency in biomechanics (\citet{Mazier_2022}). Based on a Python/C++ coding interface and the Unified Form Language, it allows to easily solve a defined variational form. Furthermore, its compatibility with open-source meshers like GMSH makes its use appealing. The project has already shown its capacity to solve large deformation problems (\citet{Mazier_2021}) and mixed formulations (\citet{Stphane2020, Urcun_2022, bulle2022}). Previous work provided the implementation of poro-mechanics within the FEniCS project (\citet{Haagenson2020,Joodat2018}). However, the FEniCS project is legacy and has been replaced by the FEniCSx project in August 2022 (\citet{ufl, basix, basix2}).

The aim of this paper is to propose a step-by-step explanation on how to implement several poro-mechanical models in FEniCSx  with a special attention to parallel computation. First, an instantaneous uni-axial confined compression of a porous elastic medium is proposed. This example corresponds to an avascular tissue. Then, the same single-compartment model is computed for a hyper-elastic solid scaffold followed by a confined bi-compartment modeling.


\section{Confined compression of a column: geometrical definition}

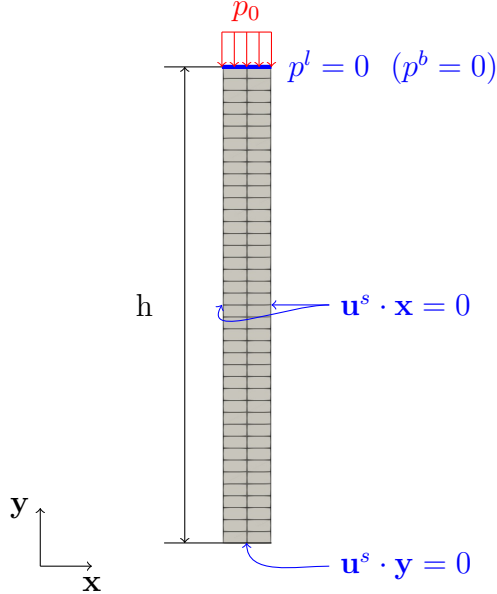
\begin{figure}[ht!]
    \centering
    \begin{tikzpicture}
        \node[anchor=south west, inner sep =0] (image) at (0,0){
    \includegraphics[width=\textwidth]{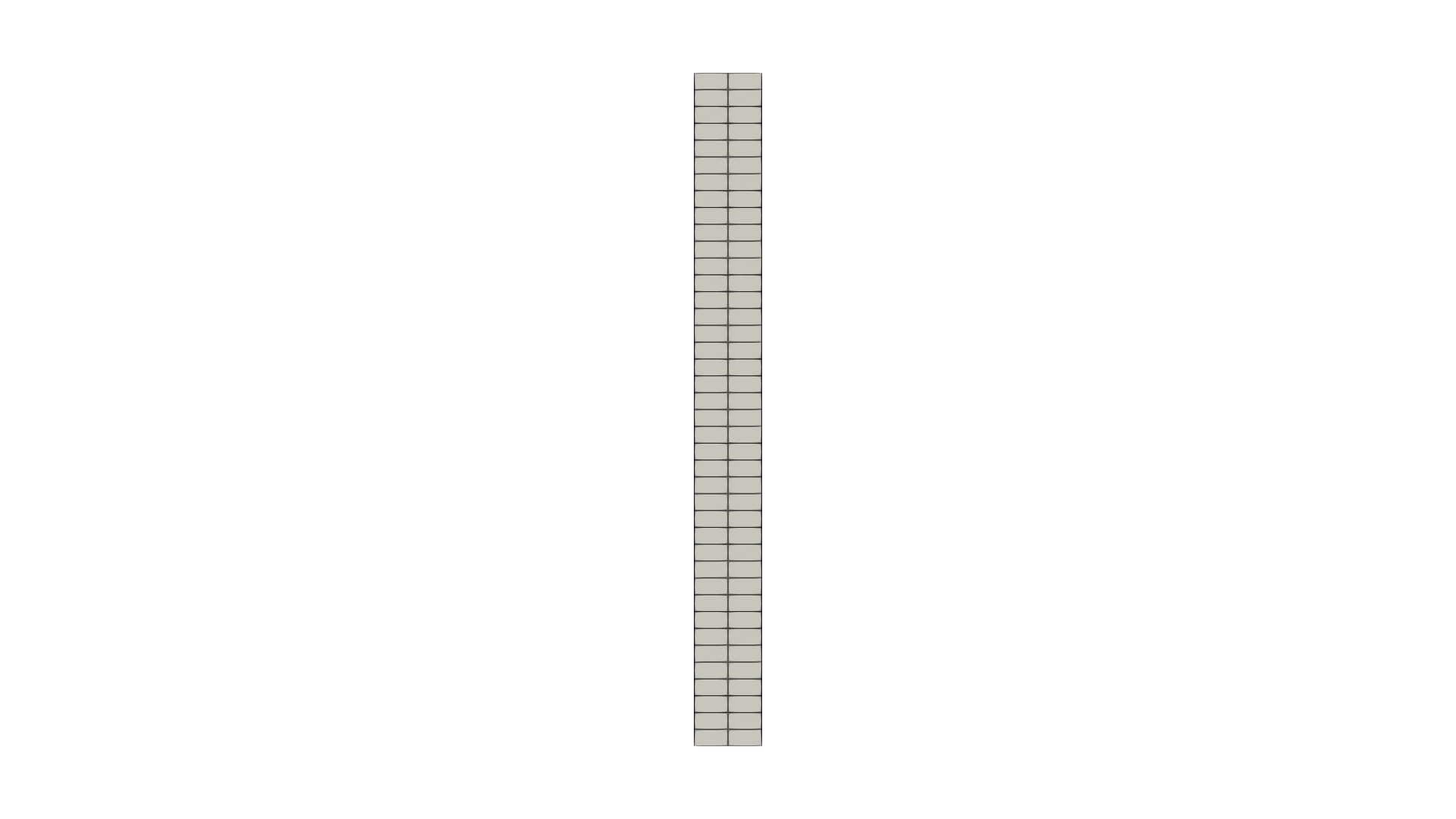}};
        \begin{scope}[x={(image.south east)},y={(image.north west)}]
            \draw[-,black] (0.42,0.91) -- (0.524,0.91);
            \draw[-,black] (0.42,0.09) -- (0.524,0.09);
            \draw[-,ultra thick,blue] (0.476,0.91) -- (0.524,0.91);
            \draw[<->,black] (0.44,0.09) -- (0.44,0.91);
            \draw[-,red] (0.476,0.97) -- (0.524,0.97);
            \draw[->,red] (0.476,0.97) -- (0.476,0.91);
            \draw[->,red] (0.524,0.97) -- (0.524,0.91);
            \draw[->,red] (0.5,0.97) -- (0.5,0.91);
            \draw[->,red] (0.512,0.97) -- (0.512,0.91);
            \draw[->,red] (0.488,0.97) -- (0.488,0.91);
            \draw[<-,blue] (0.524,0.5) to[out=0,in=180] (0.58,0.5);
            \draw[<-,blue] (0.476,0.5) to[out=-120,in=180] (0.58,0.5);
            \draw[<-,blue] (0.5,0.09) to[out=-90,in=180] (0.58,0.05);
            \draw[black] (0.42,0.5) node[left] {h};
            \draw[red] (0.5,0.97) node[above] {$p_0$};
            \draw[blue] (0.58,0.5) node[right] {$\mathbf{u}^s\cdot\mathbf{x}=0$};
            \draw[blue] (0.58,0.05) node[right] {$\mathbf{u}^s\cdot\mathbf{y}=0$};
            \draw[blue] (0.53,0.91) node[right] {$p^l=0~~(p^b=0)$};
            \draw[->,black] (0.3,0.05) -- (0.35,0.05);
            \draw[->,black] (0.3,0.05) -- (0.3,0.15);
            \draw[black] (0.3,0.15) node[left] {$\mathbf{y}$};
            \draw[black] (0.35,0.05) node[below] {$\mathbf{x}$};
        \end{scope}
    \end{tikzpicture}
    \caption{Load (red), Boundary conditions (blue) and mesh (gray) of the uni-axial confined compression of a porous 2D column of height h.}
    \label{fig:1}
\end{figure}

The time-dependent response of soft tissues are often assessed based on confined compression creep and stress relaxation test data (\citet{Budday_2019, HosseiniFarid2020, Franceschini2006, Urcun_2022}). All the benchmark examples focus on uni-axial confined compression of a column sample as shown in figure \ref{fig:1}. Both 2D and 3D geometries are studied. The column is described by its width (0.1*h) and height (h) in 2D and its length (0.1*h) in 3D. 

The dolfinx version used in this paper is v0.5.2. FEniCSx is a proficient platform for parallel computation. All described codes here-under are compatible with multi-kernel computation. The corresponding terminal command is: 

\begin{lstlisting}[language=bash]
mpirun -n <N> python3 <filename>
\end{lstlisting}

\noindent Where $<$N$>$  is the number of threads to use and $<$filename$>$ is the python code of the problem.

Within the FEniCSx software, the domain (geometry) is discretized to match with the Finite Element (FE) method. The space is thus divided in $n_x\times n_y = 2\times 40$ elements in 2D and $n_x\times n_y\times n_z= 2\times 40 \times 2$ elements in 3D. The choice of the number of elements is further discussed section \ref{linearel}. 
In this article, the meshes are directly created within the FEniCSx environment. However, as a strong compatibility exists with the GMSH API (\citet{gmsh}), it is recommended to use GMSH for this step. An example of the use of GMSH API for a more complex geometry is given section \ref{sec:4.3.1}. It is worth noting that we identify all the boundaries of interest at this step for the future declaration of boundary conditions.

\subsection{2D mesh}
Conversely to the legacy FEniCS environment, FEniCSx requires to separately import the required libraries. To create the 2D mesh, the first step is to import the following libraries:

\begin{lstlisting}[language=python]
import dolfinx
import numpy as np
from dolfinx.mesh      import create_rectangle, CellType, locate_entities, meshtags
from mpi4py            import MPI
\end{lstlisting}

Then, the domain of resolution (mesh) is computed with:

\begin{lstlisting}[language=python]
Width, Height = 1e-5, 1e-4 #[m]
nx, ny        = 2, 40      #[ ]
mesh  = create_rectangle(MPI.COMM_WORLD, np.array([[0,0],[Width, Height]]), [nx,ny], cell_type=CellType.quadrilateral)
\end{lstlisting}

Once the mesh object has been created, its boundaries are identified using couples of (marker, locator) to tag with a marker value the elements of dimension \textit{fdim} fulfilling the locator requirements.

For the 2D mesh, the (marker, locator) couples are given by:
\begin{lstlisting}[language=python]
# identifiers: 1 , 2, 3, 4 = bottom, right, top, left
boundaries = [(1, lambda x: np.isclose(x[1], 0)),
              (2, lambda x: np.isclose(x[0], Width)),
              (3, lambda x: np.isclose(x[1], Height)),
              (4, lambda x: np.isclose(x[0], 0))]
\end{lstlisting}

Finally the entities are marked by:
\begin{lstlisting}[language=python]
facet_indices, facet_markers = [], []
# dimension of the elements we are looking for
fdim = mesh.topology.dim - 1
for (marker, locator) in boundaries:
    facets = locate_entities(mesh, fdim, locator)
    facet_indices.append(facets)
    facet_markers.append(np.full_like(facets, marker))
facet_indices = np.hstack(facet_indices).astype(np.int32)
facet_markers = np.hstack(facet_markers).astype(np.int32)
sorted_facets = np.argsort(facet_indices)
# the meshtags() function requires sorted facet_indices
facet_tag = meshtags(mesh, fdim, facet_indices[sorted_facets], facet_markers[sorted_facets])
\end{lstlisting}

\subsection{3D mesh}

The method for a 3D mesh is similar to the 2D case. First, the libraries are imported and the geometry is created using a 3D function. The (marker, locator) tuples are completed to describe all the boundaries of the domain. The same tagging routine is used.

\begin{lstlisting}[language=python]
## libraries
import dolfinx
import numpy
from dolfinx.mesh      import create_box, CellType, locate_entities, meshtags
from mpi4py            import MPI
## Mesh generation
Length, Height, Width = 0.1, 1, 0.1 #[m]
nx, ny, nz = 2, 40, 2
mesh  = create_box(MPI.COMM_WORLD, numpy.array([[0.0,0.0,0.0],[Length, Height, Width]]), [nx, ny, nz], cell_type=CellType.hexahedron)
## Define the boundaries of the domain:
# 1, 2, 3, 4, 5, 6 = bottom, right, top, left, back, front
boundaries = [(1, lambda x: numpy.isclose(x[1], 0)),
			  (2, lambda x: numpy.isclose(x[0], Length)),
			  (3, lambda x: numpy.isclose(x[1], Height)),
			  (4, lambda x: numpy.isclose(x[0], 0)),
			  (5, lambda x: numpy.isclose(x[2], Width)),
			  (6, lambda x: numpy.isclose(x[2], 0))]
facet_indices, facet_markers = [], []
fdim = mesh.topology.dim - 1
for (marker, locator) in boundaries:
    facets = locate_entities(mesh, fdim, locator)
    facet_indices.append(facets)
    facet_markers.append(numpy.full_like(facets, marker))
facet_indices = numpy.hstack(facet_indices).astype(numpy.int32)
facet_markers = numpy.hstack(facet_markers).astype(numpy.int32)
sorted_facets = numpy.argsort(facet_indices)
facet_tag = meshtags(mesh, fdim, facet_indices[sorted_facets], facet_markers[sorted_facets])
\end{lstlisting}

\section{Single-compartment porous medium}
\label{sec:2}
We propose to reproduce the instantaneous uni-axial confined compression at the top surface of a single-compartment porous column of height $h$, Figure \ref{fig:1}, described by a 2D elastic or a 3D hyper-elastic solid scaffold. Regarding the 2D elastic case, the column has a height of $h=100\si{\micro\meter}$, the instantaneous load $p_0$ has a magnitude of $100\si{\pascal}$ and is applied during 6 \si{\second}. Regarding the 3D hyper-elastic case, the column has a height of $h=1\si{\meter}$, the instantaneous load $p_0$ has a magnitude of $p_0=0.3\si{\mega\pascal}$ and is applied during 100000 \si{\second}. The mechanical parameters are respectively given Table \ref{tab:1} and Table \ref{tab:2}. To assess the reliability of our results, we compare our computed solutions to Terzaghi's analytical solution and the results of \citet{SELVADURAI2016}, for the elastic and hyper-elastic scaffolds respectively.

\begin{table}[ht!]
\centering
\begin{tabular}{llll}
\hline
Parameter & Symbol & Value & Unit \\ \hline
Young modulus &  E      &   5000  &  \si{\pascal}    \\
Poisson ratio  &  $\nu$      &  0.4   &   -   \\
Intrinsic permeability &   $k^\varepsilon$     &      $1.8\times 10^{-15}$ &   \si{\square\meter}   \\
Biot coefficient   &  $\beta$      &  1     &   -   \\
Density of phase $\alpha$  &  $\rho^\alpha$      &     -  &   \si{\kilo\gram\per\cubic\meter}   \\
IF viscosity &   $\mu^l$     &      $1\times 10^{-3}$ &   \si{\pascal\second}   \\
Porosity  &   $\varepsilon^l$     &   0.5    &   -   \\
Solid grain Bulk modulus  &   $K^s $     &   $1.\times 10^{10}$    &  \si{\pascal} \\
Fluid Bulk modulus  &   $K^l $     &   $2.2\times 10^{9}$    & \si{\pascal}  \\\hline
\end{tabular}%
\caption{Elastic mechanical parameters to compare with the Terzaghi solution}
\label{tab:1}
\end{table}

\begin{table}[ht!]
\centering
\begin{tabular}{llll}
\hline
Parameter & Symbol & Value & Unit \\ \hline
Young modulus &  E      &   600000  &  \si{\pascal}    \\
Poisson ratio  &  $\nu$      &  0.3   &   -   \\
Bulk modulus  &  K      &  500000   &   \si{\pascal}   \\
Intrinsic permeability &   $k^\varepsilon$     &      $3\times 10^{-14}$ &   \si{\square\meter}   \\
IF viscosity &   $\mu^l$     &      $1\times 10^{-3}$ &   \si{\pascal\second}   \\
Porosity  &   $\varepsilon^l$     &   0.2    &   -   \\
Solid grain Bulk modulus  &   $K^s $     &   $1.\times 10^{10}$    &  \si{\pascal} \\
Fluid Bulk modulus  &   $K^l $     &   $2.2\times 10^{9}$ or $5\times 10^{5}$    & \si{\pascal} \\
Biot coefficient  &   $\beta $     &   $1-\frac{K}{K^s}\approx 1$    & -  \\\hline
\end{tabular}%
\caption{Hyper-elastic mechanical parameters from \citet{SELVADURAI2016}. In the absence of information on the porosity, solid grain bulk modulus and fluid bulk modulus, the parameter are arbitrarily chosen.}
\label{tab:2}
\end{table}

\subsection{Terzaghi's Analytical solution}

The Terzaghi consolidation problem is often used for benchmarking porous media mechanics, as an analytical solution to this problem exists. An implementation of this experiment was proposed by \citet{Haagenson2020}, within the legacy FEniCS project. The Terzaghi problem is a uni-directional confined compression experiment of a column (see Figure \ref{fig:1}). Assuming small and uni-directional strains, incompressible homogeneous phases, and constant mechanical properties, the analytical expression of the pore pressure is given in terms of series in Equation \ref{eq:30}.

\begin{align}
 p^l=\frac{4p_0}{\pi}\sum_{k=1}^{+\infty}\frac{(-1)^{k-1}}{2k-1}\cos\left[(2k-1)\frac{\pi}{2}\frac{y}{h}\right]\exp\left[-(2k-1)^2\frac{\pi^2}{4}\frac{c_vt}{h^2}\right]
 \label{eq:30}\\
 c_v = \frac{k^\varepsilon}{\mu^l( S_{\beta}+\frac{\beta^2}{M})} \label{eq:31}\\
 M = \frac{3K^s(1-\nu)}{(1+\nu)} \label{eq:32}\\
 S_{\beta} = \frac{\beta-\varepsilon^l_0}{K^s} + \frac{\varepsilon^l_0}{K^l}\label{eq:33}
\end{align}

{\noindent Where $p_0$=$\mathbf{t}^\text{imposed} \cdot  \mathbf{n}$ is the full applied load, y is the altitude, h is the initial height of the sample , $c_v$ is the consolidation coefficient defined by (Equation \ref{eq:31}), M the longitudinal modulus (Equation \ref{eq:32}), $S_{\beta}$ the inverse of the Biot Modulus (Equation \ref{eq:33}) and $\varepsilon^l_0$ is the initial porosity.}

\subsection{Governing equations}
\label{sec:2.1}
Let one consider a bi-phasic structure composed of a solid scaffold filled with interstitial fluid (IF). The medium is assumed saturated. In this section, to set up the governing equations, we make the hypothesis of a Biot coefficient equal to 1. The following convention is assumed: $\bullet^s$ denotes the solid phase and $\bullet^l$ denotes the fluid phase (IF). The primary variables of the problem are the pressure applied in the pores of the porous medium, namely $p^l$, and the displacement of the solid scaffold, namely $\mathbf{u}^s$. (Equation \ref{eq:1}) constrains the different volume fractions. The volume fraction of the phase $\alpha$ is defined by (Equation \ref{eq:2}). $\varepsilon^l$ is called the porosity of the medium.

\begin{align}
 \varepsilon^s+\varepsilon^l=1 \label{eq:1}\\
 \varepsilon^\alpha = \frac{\text{Volume}^\alpha}{\text{Volume}^{total}}
 \label{eq:2}
\end{align}

Assuming that there is no inter-phase mass transport, the continuity equations (mass conservation) of the liquid and solid phases are respectively given by Equation \ref{eq:3} and Equation \ref{eq:4}.

\begin{align}
 \frac{\partial}{\partial t}(\rho^l\varepsilon^l)+\nabla\cdot(\rho^l\varepsilon^l \mathbf{v}^l) = 0\label{eq:3}\\
 \frac{\partial}{\partial t}(\rho^s(1-\varepsilon^l))+\nabla\cdot(\rho^s(1-\varepsilon^l)\mathbf{v^s}) = 0 
 \label{eq:4}
\end{align}

Regarding the distributivity of the divergence term, with \textit{a} scalar and \textbf{V} vector,

\begin{equation}
    \nabla\cdot(a\mathbf{V}) = a \nabla\cdot(\mathbf{V}) + \nabla a\cdot\mathbf{V}
    \label{eq:5}
\end{equation}

Applied to \ref{eq:3} and Equation \ref{eq:4}, and considering the definition of the material derivative, $\frac{\mathrm{D}^s}{\mathrm{D}t} f = \frac{\partial f}{\partial t} + \mathbf{\nabla} f \cdot \mathbf{v}^s $, the continuity equations are given by:

\begin{align}
 \frac{\mathrm{D}^s}{\mathrm{D}t}(\rho^s(1-\varepsilon^l))+\rho^s(1-\varepsilon^l)\nabla\cdot\mathbf{v^s} = 0 \label{eq:6}\\
 \frac{\mathrm{D}^s}{\mathrm{D}t}(\rho^l\varepsilon^l)+\nabla\cdot(\rho^l\varepsilon^l (\mathbf{v}^l-\mathbf{v}^s)) + \rho^l\varepsilon^l \nabla\cdot\mathbf{v^s} = 0
 \label{eq:7}
\end{align}

For the fluid phase, the Darcy's law (Equation \ref{eq:8}) is used to evaluate the fluid flow in the porous medium. 

\begin{align}
 \varepsilon^l(\mathbf{v}^l-\mathbf{v}^s) = - \frac{k^\varepsilon}{\mu^l}(\mathbf{\nabla} p^l-\rho^l\mathbf{g})
 \label{eq:8}
\end{align}

{\noindent Where $k^\varepsilon$ is the intrinsic permeability (\si{\square \meter}), $\mu^l$ is the dynamic viscosity (\si{\pascal \second}) and $\mathbf{g}$ the gravity.}

Introducing the state law $\frac{1}{\rho^\alpha}\frac{\mathrm{D}^s\rho^\alpha}{\mathrm{D}t}=\frac{1}{K^\alpha}\frac{\mathrm{D}p^\alpha}{\mathrm{D}t}$, $K^\alpha$ being the bulk modulus of the phase alpha, the Darcy's law and summing \ref{eq:6} and Equation \ref{eq:7}, we obtain:

\begin{equation}
    \left(\frac{\varepsilon^l}{K^l}+\frac{1-\varepsilon^l}{K^s} \right)\frac{\mathrm{D}^s p^l}{\mathrm{D}t}+\nabla\cdot\mathbf{v}^s -\nabla\cdot\left(\frac{k^\varepsilon}{\mu^l}\mathbf{\nabla}p^l\right) = 0
    \label{eq:9}
\end{equation}

{\noindent Where $S= \left(\frac{\varepsilon^l}{K^l}+\frac{1-\varepsilon^l}{K^s} \right)$ is called the storativity coefficient.}

Once the continuity equations are settled, one can define the quasi-static momentum balance of the porous medium, Equation \ref{eq:10}.

\begin{equation}
    \mathbf{\nabla}\cdot\mathbf{t}^{\text{tot}} = 0
    \label{eq:10}
\end{equation}

\noindent Where $\mathbf{t}^{\text{tot}}$ is the total Cauchy stress tensor. We introduce an effective stress tensor denoted $\mathbf{t}^\text{eff}$, responsible for all deformation of the solid scaffold. Then, $\mathbf{t}^{\text{tot}}$ can be expressed as:

\begin{equation}
    \mathbf{t}^{\text{tot}}=\mathbf{t}^{\text{eff}}-\beta p^l\mathbf{I_d}
\end{equation}
{\noindent Where $\mathbf{I_d}$ is the identity matrix and $\beta$ is the Biot coefficient.}

Finally, the governing equations of this single compartment porous medium are:

\begin{align}
    \left(\frac{\varepsilon^l}{K^l}+\frac{1-\varepsilon^l}{K^s} \right)\frac{\mathrm{D}^s p^l}{\mathrm{D}t}+\nabla\cdot\mathbf{v}^s -\nabla\cdot\left(\frac{k^\varepsilon}{\mu^l}\mathbf{\nabla}p^l\right) = 0~\text{on }\Omega\label{eq:12}\\
  \mathbf{\nabla}\cdot\mathbf{t}^{\text{tot}} = 0~\text{on }\Omega
 \label{eq:13}
\end{align}

Three boundaries are defined: the first one, $\Gamma_u$ has imposed displacement (Equation \ref{eq:14}), the second one $\Gamma_s$ has imposed external forces (Equation \ref{eq:15}) and $\Gamma_p$ is submitted to an imposed pressure (fluid leakage condition (Equation \ref{eq:16})). We obtain:

\begin{align}
 \mathbf{t}^\text{eff} = \mathbf{t}^\text{imposed}~\text{on}~\Gamma_s \label{eq:14}\\
 \mathbf{u}^s=\mathbf{u}^\text{imposed}~\text{on}~\Gamma_u \label{eq:15}\\
 p=0~\text{on}~\Gamma_p \label{eq:16}
\end{align}

According to Figure \ref{fig:1}, $\Gamma_p = \Gamma_s$ is the top surface and $\Gamma_u$ covers the lateral and bottom surfaces.

\subsection{Effective stress}
Two type of solid constitutive laws are considered: an elastic scaffold and a hyper-elastic one.

\subsubsection{Linear elasticity}
In case of an elastic scaffold, the effective stress tensor is defined as follows:

\begin{align}
     \mathbf{\epsilon}(\mathbf{u})=\frac{1}{2}(\nabla\mathbf{u}+\nabla\mathbf{u}^\text{T})\label{eq:19}\\
     \mathbf{t}^\text{eff} = 2\mu\mathbf{\epsilon}(\mathbf{u}^s)+\lambda \text{tr}(\mathbf{\epsilon}(\mathbf{u}^s))\mathbf{I_d} \label{eq:20}
\end{align}
{\noindent Where $\mathbf{I_d}$ is the identity matrix and ($\lambda,\mu$) the Lame coefficients.}

\subsubsection{Hyper-elasticity}
In case of a hyper-elastic scaffold, other quantities are required.
Let one introduce the deformation gradient $\mathbf{F}$:
 
 \begin{equation}
     \mathbf{F}=\mathbf{I}_d+\mathbf{\nabla}\mathbf{u}^s
     \label{eq:21}
 \end{equation}
 
Then, J is the determinant of $\mathbf{F}$:
 
 \begin{align}
     J = \text{det}(\mathbf{F})
     \label{eq:22}
 \end{align}

 According to the classic formulation of a finite element procedure, we introduce $\mathbf{C}$ the right Cauchy-Green stress tensor and its first invariant $I_1$. By definition:
 
 \begin{align}
     \mathbf{C} = \mathbf{F}^{\text{T}}\mathbf{F} \label{eq:23}\\
     I_1 = \text{Tr}(\mathbf{C})
     \label{eq:24}
 \end{align}

The theory of hyper-elasticity defines a potential of elastic energy $W(\mathbf{F})$. The generalized Neo-Hookean potential (Equation \ref{eq:26b}) introduced by \citet{treloar1975physics}, implemented in Abaqus and used by \citet{SELVADURAI2016} is evaluated in this article.
\begin{equation}
    {W}(\mathbf{F})= \frac{\mu}{2}(\mathrm{J}^{-2/3}I_1-\text{tr}(\mathbf{I_d}))+\left(\frac{\lambda}{2}+\frac{\mu}{3}\right)*(\mathrm{J}-1)^2
    \label{eq:26b}
\end{equation}

However, other potential were developed. It was shown that the hyper-elastic potential can be expressed as the combination of a isochoric component and a volumetric component (\citet{ SIMO1988,Horgan2004,Michele2018}). We define the lame coefficients by $\mu=\frac{E}{2(1-\nu)}$ and $\lambda=\frac{E\nu}{(1+\nu)(1-2\nu)}$. For a Neo-Hookean material, we further have:

\begin{equation}
    W(\mathbf{F})=\Tilde{W}(I_1,J)+U(J)
    \label{eq:25}
\end{equation}

Where $\Tilde{W}(I_1,J)$ is the isochoric part and $U(J)$ the volumetric one. The study of \citet{SELVADURAI2016} presented a compressible case ($\nu=0.3$) reaching high deformation. Therefore, a compressible formulation of the Neo-Hookean strain-energy potential from \citet{Pence2014,Horgan2004} is also computed for comparison. Therefore, the implemented isochoric part of the strain energy potential is:  

\begin{equation}
    \Tilde{W}_1(I_1,J)= \frac{\mu}{2}(I_1-\text{tr}(\mathbf{I_d})-2\log[\mathrm{J}])
    \label{eq:26}
\end{equation}

Two different volumetric parts ($U_1$ and $U_2$) which were proposed in \citet{doll2000development} are implemented, 

\begin{equation}
    U_1(J)=\frac{\lambda}{2} \log[\mathrm{J}]^2
    \label{eq:27}
\end{equation}
\begin{equation}
    U_2(J)=\frac{\lambda}{2} (J-1)^2
    \label{eq:28}
\end{equation}

Finally, from the potential (Equation \ref{eq:25} or \ref{eq:26b}) derives the first Piola-Kirchhoff stress tensor as the effective stress such that:

\begin{equation}
    \mathbf{t}^\text{eff}=\frac{\partial W}{\partial \mathbf{F}}
    \label{eq:29}
\end{equation}

\subsection{Variational formulation}

For the computation of the Finite Element (FE) model, the variational form of Equation \ref{eq:12} and Equation \ref{eq:13} is introduced. Let one consider (q,v) the test functions defined in the mixed space $\text{L}_0^2(\Omega)\times[\text{H}^1(\Omega)]^2$. 

With a first order approximation in time, Equation \ref{eq:12} gives:

\begin{align}
    \begin{split}  
    \frac{S}{dt}\int_{\Omega} (p^l-p^l_n)q\text{d}\Omega+\frac{1}{dt}\int_{\Omega} \mathbf{\nabla}\cdot(\mathbf{u}^s-\mathbf{u}^s_n)q\text{d}\Omega \\
    + \frac{k^\varepsilon}{\mu^l} \int_{\Omega} \mathbf{\nabla}p^l\mathbf{\nabla}q\text{d}\Omega = 0, \forall~q\in~\text{L}_0^2(\Omega)
    \end{split}
    \label{eq:17}
\end{align}

Similarly, by integrating by part Equation \ref{eq:13}, and including the Neumann boundary conditions, we get:

\begin{align}
\begin{split}
        \int_{\Omega} \mathbf{t}^\text{eff}:\mathbf{\nabla}\mathbf{v}\text{d}\Omega-\int_{\Omega}\beta p^l\mathbf{\nabla}\cdot\mathbf{v}\text{d}\Omega - \int_{\Gamma_s} \mathbf{t}^\text{imposed} \cdot  \mathbf{n} \cdot \mathbf{v} \text{d}\Gamma_s=0,\\ \forall~\mathbf{v}\in~[\text{H}^1(\Omega)]^2
    \label{eq:18}
\end{split}
\end{align}

The first order approximation in time impose to define the initial conditions which are fixed according to Table \ref{tab:3}.

\begin{table}[ht!]
\centering
\begin{tabular}{llll}
\hline
Parameter & Symbol & Value & Unit \\ \hline
     Displacement     &   $\mathbf{u}^s$     &  0     &   \si{\meter}   \\
     Displacement at previous step     &   $\mathbf{u}^s_n$     &  0     &   \si{\meter}   \\
     IF pressure    &     $p^l$    &  $\mathbf{t}^\text{imposed} \cdot  \mathbf{n}$     &     \si{\pascal}  \\
     IF pressure at previous step   &     $p^l_n$    &  0     &     \si{\pascal}  \\ \hline
\end{tabular}%
\caption{Initial conditions for the single compartment model}
\label{tab:3}
\end{table}

Finally, the problem to solve is: $\text{Find}~(p^l, \mathbf{u}^s)\in\text{L}_0^2(\Omega)\times[\text{H}^1(\Omega)]^2$ such that Equation \ref{eq:17} and Equation \ref{eq:18} are verified.


\subsection{2D linear elastic solid scaffold}

\subsubsection{FEniCSx implementation}
This section aims to provide a possible implementation of a 2D elastic problem and its comparison with the Terzaghi analytical solution. Conversely to the former FEniCS project, Dolfinx is based on a more explicit use of the libraries and requires to import them in the FEniCSx environment separately.
Therefore, each function used in the following implementation of the problem needs to be imported as a first step.

\begin{lstlisting}[language=python]
import numpy as np
from dolfinx           import nls
from dolfinx.fem.petsc import NonlinearProblem
from ufl               import VectorElement, FiniteElement, MixedElement, TestFunctions, TrialFunction
from ufl               import Measure, FacetNormal
from ufl               import nabla_div, dx, dot, inner, grad, derivative, split
from petsc4py.PETSc    import ScalarType
from mpi4py            import MPI
from dolfinx.fem       import (Constant, dirichletbc, Function, FunctionSpace, locate_dofs_topological)
from dolfinx.io        import XDMFFile
\end{lstlisting}

Then, the time parametrization is introduced, the load value T such that $\mathbf{t}^\text{imposed} =  p_0 \cdot \mathbf{n}$ with $\mathbf{n}$ the outward normal to the mesh, and the material parameters which are defined as ufl constants over the mesh.

\begin{lstlisting}[language=python]
## Time parametrization
t         = 0                # Start time
Tf        = 6.               # End time
num_steps = 1000             # Number of time steps
dt        = (Tf-t)/num_steps # Time step size
## Material parameters
E            = Constant(mesh, ScalarType(5000))  
nu           = Constant(mesh, ScalarType(0.4))
lambda_m     = Constant(mesh, ScalarType(E.value*nu.value/((1+nu.value)*(1-2*nu.value))))
mu           = Constant(mesh, ScalarType(E.value/(2*(1+nu.value))))
rhos         = Constant(mesh, ScalarType(1))
kepsilon     = Constant(mesh, ScalarType(1.8e-15)) 
mul          = Constant(mesh, ScalarType(1e-2))  
rhol         = Constant(mesh, ScalarType(1))
beta         = Constant(mesh, ScalarType(1))
epsilonl     = Constant(mesh, ScalarType(0.2))
Kf           = Constant(mesh, ScalarType(2.2e9))
Ks           = Constant(mesh, ScalarType(1e10))
S            = (epsilonl/Kf)+(1-epsilonl)/Ks
## Mechanical loading 
pinit = 100 #[Pa]
T     = Constant(mesh,ScalarType(-pinit))
\end{lstlisting}

The surface element for integration based on the tags and the normals of the mesh are computed. 
\begin{lstlisting}[language=python]
# Create the surfacic element
ds = Measure("ds", domain=mesh, subdomain_data=facet_tag)
# compute the mesh normals to express t^imposed = T.normal
normal = FacetNormal(mesh)
\end{lstlisting}

Two type of elements are defined for displacement and pressure, then combined to obtain the mixed space (MS) of the solution.
\begin{lstlisting}[language=python]
displacement_element  = VectorElement("CG", mesh.ufl_cell(), 2)
pressure_element      = FiniteElement("CG", mesh.ufl_cell(), 1)
MS                    = FunctionSpace(mesh, MixedElement([displacement_element,pressure_element]))
\end{lstlisting}

The space of resolution being defined, we can introduce the Dirichlet boundary conditions according to Equation \ref{eq:15}, Equation \ref{eq:16} and Figure \ref{fig:1}.

\begin{lstlisting}[language=python]
# 1 = bottom: uy=0, 2 = right: ux=0, 3=top: pl=0 drainage, 4=left: ux=0
bcs    = []
fdim = mesh.topology.dim - 1
# uy=0
facets = facet_tag.find(1)
dofs   = locate_dofs_topological(MS.sub(0).sub(1), fdim, facets)
bcs.append(dirichletbc(ScalarType(0), dofs, MS.sub(0).sub(1)))
# ux=0
facets = facet_tag.find(2)
dofs   = locate_dofs_topological(MS.sub(0).sub(0), fdim, facets)
bcs.append(dirichletbc(ScalarType(0), dofs, MS.sub(0).sub(0)))
# ux=0
facets = facet_tag.find(4)
dofs   = locate_dofs_topological(MS.sub(0).sub(0), fdim, facets)
bcs.append(dirichletbc(ScalarType(0), dofs, MS.sub(0).sub(0)))
# leakage p=0
facets = facet_tag.find(3)
dofs   = locate_dofs_topological(MS.sub(1), fdim, facets)
bcs.append(dirichletbc(ScalarType(0), dofs, MS.sub(1)))
\end{lstlisting}

The problem depends on the time Equation \ref{eq:17}. Initial conditions in displacement and pressure are required. Therefore, we defined X0 the unknown function and Xn the solution at the previous step. Giving the \textit{collapse}() function, the initial displacement function Un\_ and its mapping within the Xn solution are identified. Then, its values are set to 0 and reassigned in Xn using the map. Xn.x.\textit{scatter\_forward}() allows to update the values of Xn in case of parallel computation. The same method is used to set up the initial pressure field. To fit with the studied problems, the load is instantaneously applied. Therefore, the initial pore pressure of the sample is assumed equal to $p_0$. 

\begin{lstlisting}[language=python]
# X0, Xn: Solution and previous functions of space
X0 = Function(MS)
Xn = Function(MS)
# Initial values
# Solid Displacement
Un_, Un_to_MS = MS.sub(0).collapse()
FUn_ = Function(Un_)
with FUn_.vector.localForm() as initial_local:
    initial_local.set(ScalarType(0.0)) 
# Assign in Xn and broadcast to all the threads
Xn.x.array[Un_to_MS] = FUn_.x.array
Xn.x.scatter_forward()
# IF Pressure
Pn_, Pn_to_MS = MS.sub(1).collapse()
FPn_ = Function(Pn_)
with FPn_.vector.localForm() as initial_local:
    initial_local.set(ScalarType(pinit)) 
# Assign in Xn and broadcast to all the threads
Xn.x.array[Pn_to_MS] = FPn_.x.array
Xn.x.scatter_forward()
\end{lstlisting}

The deformation and effective stress given Equation \ref{eq:19} and Equation \ref{eq:20} are defined by the following function:

\begin{lstlisting}[language=python]
def teff_Elastic(u,lambda_m,mu):
    from ufl import sym, grad, nabla_div, Identity
    ## Deformation
    epsilon = sym(grad(u))
    ## Stress
    return lambda_m * nabla_div(u) * Identity(u.geometric_dimension()) + 2*mu*epsilon
\end{lstlisting}

Finally, splitting the two functions X0, Xn, and introducing the test functions, the weak form is implemented as follows.

\begin{lstlisting}[language=python]
u,p    =split(X0)
u_n,p_n=split(Xn)
# Set up the test functions
v,q = TestFunctions(MS)
# Equation 33
F  = (1/dt)*nabla_div(u-u_n)*q*dx + (kepsilon/mul)*dot(grad(p),grad(q))*dx  + ( S/dt )*(p-p_n)*q*dx
# Equation 34
F += inner(grad(v),teff(u))*dx - beta * p * nabla_div(v)*dx - T*inner(v,normal)*ds(3)
\end{lstlisting}

Introducing the trial function of the mixed space dX0, we define the non-linear problem based on the variational form, the unknown, the boundary conditions and the Jacobian:
\begin{lstlisting}[language=python]
dX0     = TrialFunction(MS)
Js      = derivative(F, X0, dX0)
Problem = NonlinearProblem(F, X0, bcs = bcs, J = Js)
\end{lstlisting}

\subsubsection{Solving and results}
\label{linearel}
To solve the non-linear problem defined here-above, a Newton solver is tuned.

\begin{lstlisting}[language=python]
solver  = nls.petsc.NewtonSolver(mesh.comm, Problem)
# Absolute tolerance
solver.atol = 5e-10
# relative tolerance
solver.rtol = 1e-11
solver.convergence_criterion = "incremental"
\end{lstlisting}

The parameters were set according to Table \ref{tab:1}. During the resolution, we computed for each step the error in $L^2$-norm in pressure defined Equation \ref{eq:34}. These formulations are easily evaluated within the FEniCSx environment by defining the following functions:

\begin{equation}
    E({p^l})~=~\frac{\sqrt{\int_\Omega (p^l-p^{ex})^2\mathrm{d} x}}{\sqrt{\int_\Omega (p^{ex})^2\mathrm{d} x}}
    \label{eq:34}
\end{equation}

Where $p^{ex}$ is the exact solution, computed from the Terzaghi's analytical formula.

\begin{lstlisting}[language=python]
def terzaghi_p(x):
    kmax=1e3
    p0,L=pinit,Height
    cv = kepsilon.value/mul.value*(lambda_m.value+2*mu.value)
    pression=0
    for k in range(1,int(kmax)):
        pression+=p0*4/np.pi*(-1)**(k-1)/(2*k-1)*np.cos((2*k-1)*0.5*np.pi*(x[1]/L))*np.exp(-(2*k-1)**2*0.25*np.pi**2*cv*t/L**2)
    pl=pression
    return pl
def L2_error_p(mesh,pressure_element,__p):
    V2 = FunctionSpace(mesh, pressure_element)
    pex = Function(V2)
    pex.interpolate(terzaghi_p)
    L2_errorp, L2_normp = form(inner(__p - pex, __p - pex) * dx), form(inner(pex, pex) * dx)
    error_localp = assemble_scalar(L2_errorp)/assemble_scalar(L2_normp)
    error_L2p = np.sqrt(mesh.comm.allreduce(error_localp, op=MPI.SUM))
    return error_L2p
\end{lstlisting}

To get a code suitable for parallel computation, the solutions needed to be gathered on a same processor using the MPI.\textit{allreduce}() function.
Once the error functions were defined, the problem is solved within the time loop:

\begin{lstlisting}[language=python]
# Create an output xdmf file to store the values
xdmf = XDMFFile(mesh.comm, "./terzaghi.xdmf", "w")
xdmf.write_mesh(mesh)
# Solve the problem and evaluate values of interest
t = 0
L2_p = np.zeros(num_steps, dtype=PETSc.ScalarType)
for n in range(num_steps):
    t += dt
    num_its, converged = solver.solve(X0)
    X0.x.scatter_forward()
    # Update Value
    Xn.x.array[:] = X0.x.array
    Xn.x.scatter_forward()
    __u, __p = X0.split()
    # Export the results
    __u.name = "Displacement"
    __p.name = "Pressure"
    xdmf.write_function(__u,t)
    xdmf.write_function(__p,t)
    # Compute L2 norm for pressure
    error_L2p     = L2_error_p(mesh,pressure_element,__p)
    L2_p[n] = error_L2p
    # Solve tracking
    if mesh.comm.rank == 0:
        print(f"Time step {n}, Number of iterations {num_its}, Load {T.value}, L2-error p {error_L2p:.2e}")    
xdmf.close()
\end{lstlisting}

\begin{figure}[ht!]
\centering
    \begin{subfigure}[t]{0.47\textwidth}
        \includegraphics[width=\textwidth]{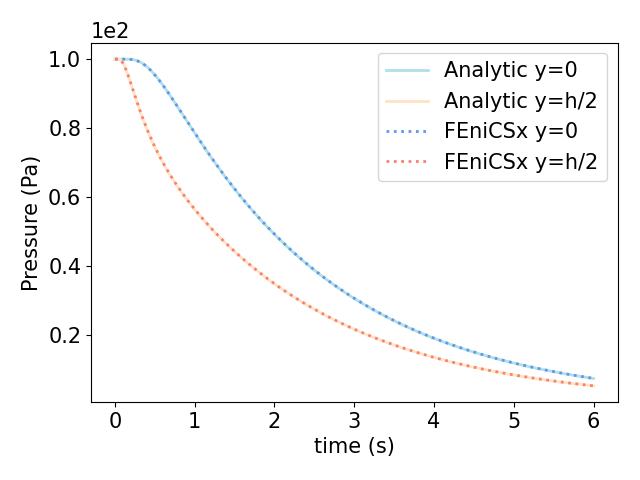}
        \caption{}
        \label{fig:2a}
    \end{subfigure}
    \begin{subfigure}[t]{0.47\textwidth}
        \includegraphics[width=\textwidth]{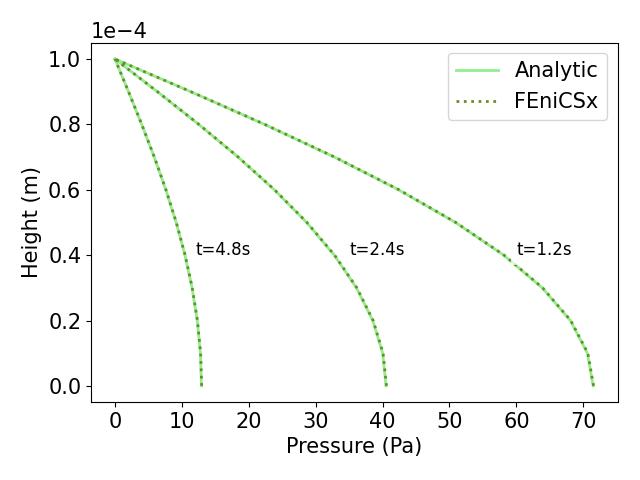}
        \caption{}
        \label{fig:2b}
    \end{subfigure}   
\caption{Comparison of the computed pore pressure against the analytical solution: in (a) time and (b) space. The pressure was well recovered based on the evaluation of the $L^2$-norm error $(3.57\pm2.46)\times10^{-3}$. }
\label{fig:2}
\end{figure}

The results obtained for pressure and displacements are provided Figure \ref{fig:2}. The code to evaluate the pressure at given points is provided \ref{appendix:eval}.

The curves show the efficiency of the simulation to reproduce the analytical solution. The accuracy of the simulation was also supported by the estimation of the error based on the $L^2$-norm of the pressure equal to $(3.57\pm2.46)\times10^{-3}$ which is deemed satisfactory. The same problem was solved using the legacy FEniCS version. The proposed FEniCSx implementation was faster. It was computed in 9.48 seconds compared to the previously 31.82 seconds.

To show the efficiency of the parallel computation, the 3D case \ref{appendix:3D:case} is considered. For a given spatio-temporal discretization, a larger computational time of 1 hour 4 minutes 29 seconds is needed using FEniCSx. To reduce the time, the code naturally supports parallel computation. The same code was run for several number of threads. Computed on 2 threads, the code required 53 minutes 27 seconds. For 4 threads, the running time was further reduced to 46 minutes 27 seconds. Finally, using 8 threads, the computation time was reduced up to 28 minutes 9 seconds.

\begin{figure}[ht!]
    \centering
    \begin{subfigure}[t]{0.47\textwidth}
        \includegraphics[width=\textwidth]{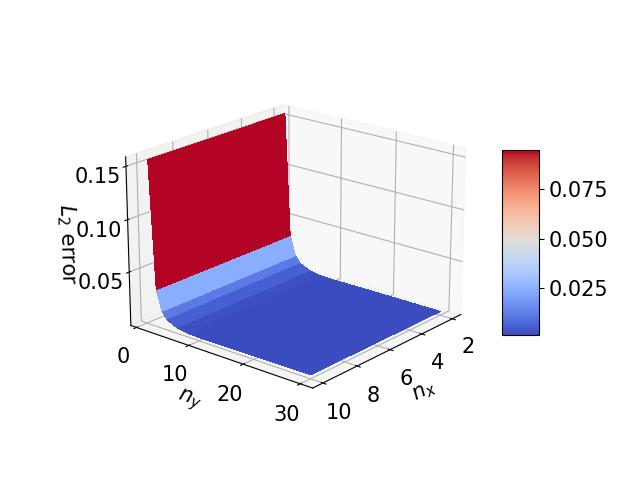}
        \caption{}
    \end{subfigure}
    \begin{subfigure}[t]{0.47\textwidth}
        \includegraphics[width=\textwidth]{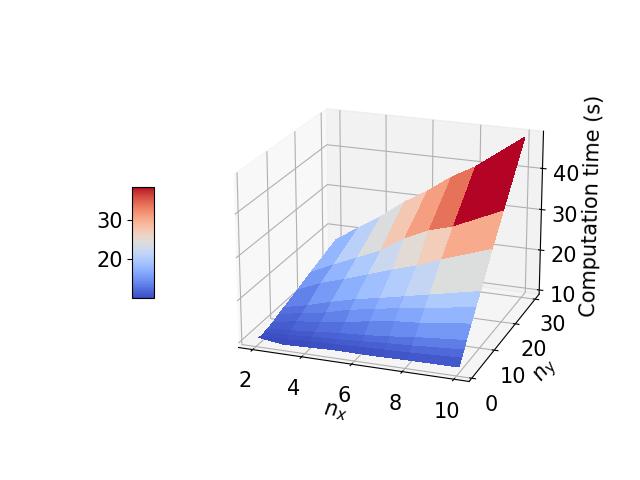}
        \caption{}
    \end{subfigure}
    \caption{Convergence analysis for a $n_x\times n_y$ discretized mesh: L-2 norm (a) and Computation time (b)} 
    \label{fig:3}
\end{figure}

Finally, a convergence analysis on the meshing of the column was carried out. The $L_2$ error metric was used and its evolution for a $n_x\times n_y$ discretized mesh is given Figure \ref{fig:3}. As we could have expected from the 1D behavior of a confined compression Terzaghi case, the error is almost independent from the $n_x$ choice. Figure \ref{fig:3}(a) shows that a $n_y\geq10$ gives better estimations. According to Figure \ref{fig:3}(b), a balance between precision and computation time must be considered. The more elements, the higher the computation time. To ensure obtaining a reliable solution, a mesh of $n_x\times n_y = 2 \times 40$ was used.

\subsection{3D hyper-elastic scaffold}

\subsubsection{FEniCSx implementation}

The implementation method of the 3D case is the same. However, special attention must be placed on the boundary. Indeed, moving from 2D to 3D introduces two more boundaries. Therefore, the Dirichlet boundary conditions definition is completed with:

\begin{lstlisting}[language=python]
# uz=0
facets = facet_tag.find(5)
dofs   = locate_dofs_topological(MS.sub(0).sub(2), fdim, facets)
bcs.append(dirichletbc(ScalarType(0), dofs, MS.sub(0).sub(2)))
# uz=0
facets = facet_tag.find(6)
dofs   = locate_dofs_topological(MS.sub(0).sub(2), fdim, facets)
bcs.append(dirichletbc(ScalarType(0), dofs, MS.sub(0).sub(2)))
\end{lstlisting}

The effective stress tensor is also different.  As an example, the stress tensor resulting from the potential $W(\mathbf{F})=\Tilde{W}_1(I_1,J)+U_1(J)$ is defined in FEniCSx by:

\begin{lstlisting}[language=python]
def teff(u,lambda_m,mu):
    from ufl import variable, Identity, grad, det, tr, ln, diff
    ## Deformation gradient
    F = variable(Identity(len(u)) + grad(u))
    J  = variable(det(F))
    ## Right Cauchy-Green tensor
    C = variable(F.T * F)
    ##Invariants of deformation tensors
    Ic = variable(tr(C))
    ## Potential
    W = (mu / 2) * (Ic - 3) - mu * ln(J) + (lambda_m / 2) * (ln(J))**2
    return diff(W, F)
\end{lstlisting}

All other developed potential are available in the supplementary material.

\subsubsection{Results}
The same solver options as for the 2D case were used. To limit the computation time, the time step was made variable: dt=500 for $t\in[0,20000]$, dt=1000 for $t\in[20000,60000]$ and dt=10000 for $t\in[60000,100000]$. A total of 84 time steps was then considered. 

\begin{figure}[H]
\centering
   \begin{subfigure}[t]{0.47\textwidth}
        \includegraphics[width=\textwidth]{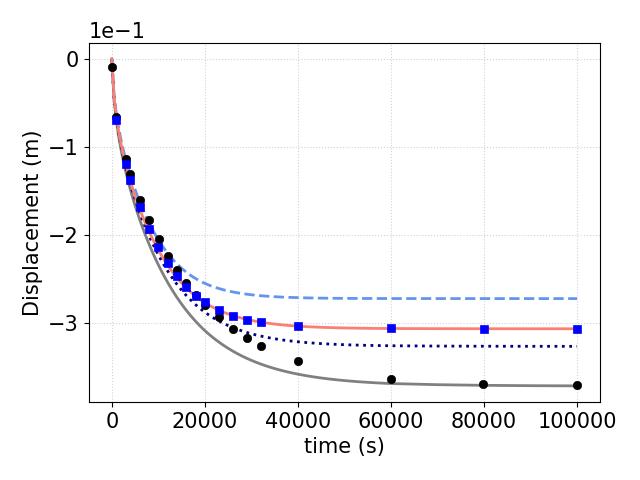}
        \caption{}
        \label{fig:4a}
    \end{subfigure}
    \begin{subfigure}[t]{0.47\textwidth}
        \includegraphics[width=\textwidth]{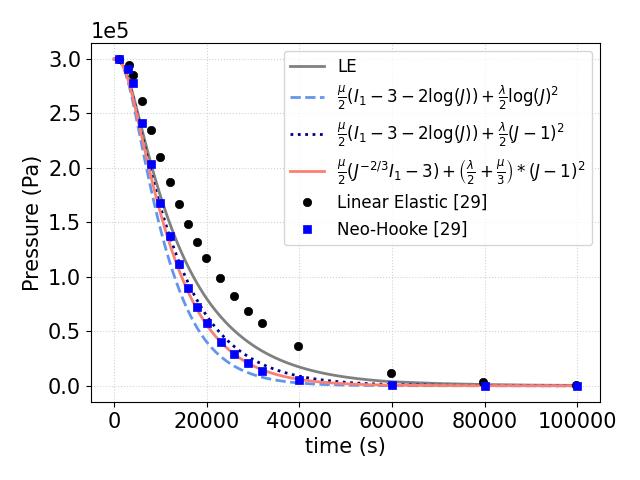}
        \caption{}
        \label{fig:4b}
    \end{subfigure} 
    \begin{subfigure}[t]{0.47\textwidth}
        \includegraphics[width=\textwidth]{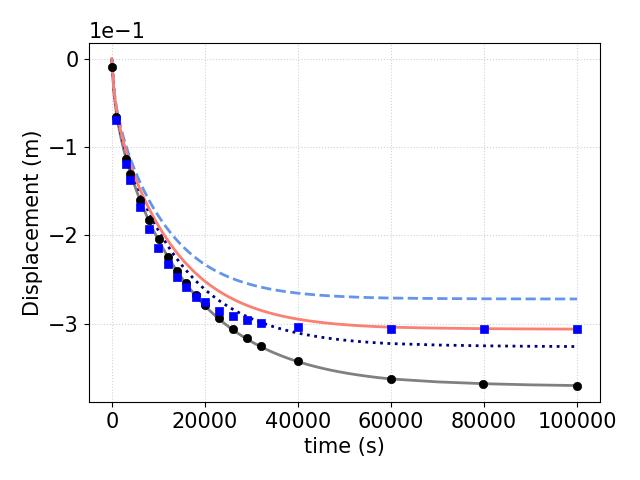}
        \caption{}
        \label{fig:4c}
    \end{subfigure}
    \begin{subfigure}[t]{0.47\textwidth}
        \includegraphics[width=\textwidth]{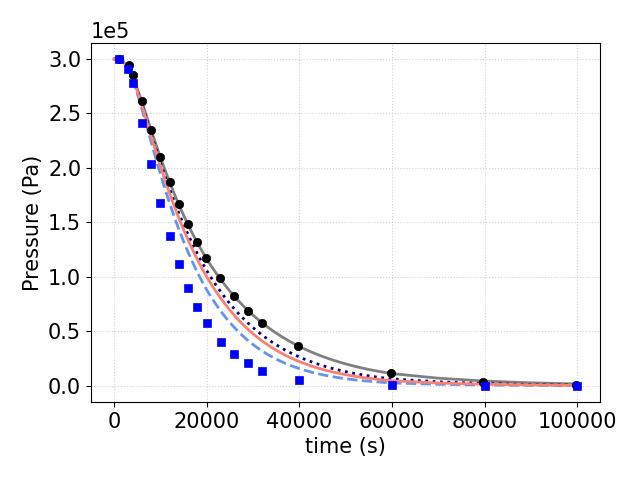}
        \caption{}
        \label{fig:4d}
    \end{subfigure}    
\caption{$K^f=2.2\times10^{9}$: (a) Displacement of the to surface points and (b) pressure at the bottom of the column. $K^f=5\times10^{5}$: (c) Displacement of the to surface points and (d) pressure at the bottom of the column. The computed Linear Elastic (LE) and Neo-Hookean (NH) for both volumetric functions and the found calibrated parameters are super-imposed with the expected values from \citet{SELVADURAI2016}.}
\label{fig:4}
\end{figure}

The parameters were set according to Table \ref{tab:2}. The results for the previously defined strain-energy potential are given Figure \ref{fig:4}. Each finite element problem was computed in $23.6\pm4.3$ seconds on 8 threads. Independently from the choice of the potential, the consolidated pressure was retrieved. On the contrary, the resulting displacement depends on the chosen potential but a same order of magnitude is found for all the cases and describe well the observations proposed in \citet{SELVADURAI2016}. 

In the absence of information about the porosity or the fluid bulk modulus in the referent study, two fluid bulk modulus were considered. 
In case where the fluid bulk modulus is made close to the water one ($K^f=2.2\times10^{9}$), the hyper-elastic material well recovers the expected values. However, mismatches appear for a linear scaffold. This can result from the use of a elastic law for large deformations. {For example, an Abaqus user would specify the geometric non-linearity whereas no special care was applied in the current example.} Furthermore, in case of a lower value of the fluid bulk modulus $K^f=5\times10^{5}$ (\textit{i.e.,} it can correspond to a non-constant value of the permeability and the porosity), the elastic behavior was recovered but differences on the hyper-elastic formulation were obtained. 

{We believe that these differences result from a permeability depending on the stress state of the column which has not been developed in the referent paper ('Initial values of the permeability and viscosity are the same for all three materials.' from \citet{SELVADURAI2016}).}

\section{Confined bi-compartment porous-elastic medium}
\label{sec:4}

Sections \ref{sec:2} proposed a poro-mechanical modeling of a single-compartment porous medium (suitable for an avascularised tissue for instance). In case of \textit{in vivo} modeling, at least one more fluid phase is required: the blood. A 3D confined compression example of a column of height 100 \si{\micro\meter} is proposed, based on the here-after variational formulation and \citet{Scium2021} study. The load is applied as a sinusoidal ramp up to the magnitude of 100 \si{\pascal} during 5 seconds. Then, the load is sustained for 125 seconds.

For more complex geometries, a gmsh example of a rectangle geometry indented by a cylindrical beam on its top surface and the corresponding local refinement are proposed \ref{sec:4.3.1}.

\begin{table}[ht!]
\centering
\begin{tabular}{llll}
\hline
Parameter & Symbol & Value & Unit \\ \hline
Young modulus &  E      &   5000  &  \si{\pascal}    \\
Poisson ratio  &  $\nu$      &  0.2   &   -   \\
IF viscosity &   $\mu^l$     &      1 &   \si{\pascal\second}   \\
Intrinsic permeability &   $k^{\varepsilon}$     &      $1.\times 10^{-14}$ &   \si{\square\meter}   \\
Biot coefficient   &  $\beta$      &  1     &   -   \\
Density of phase $\alpha$  &  $\rho^\alpha$      &     -  &   \si{\kilo\gram\per\cubic\meter}   \\
Porosity  &   $\varepsilon^l$     &   0.5    &   -   \\
Vessel Bulk modulus  &   $K^\nu $     &   $1\times 10^{3}$    & \si{\pascal}  \\
vessel Intrinsic permeability &   $k^\varepsilon_b$     &      $2\times 10^{-16}$ or  $4\times 10^{-16}$ &   \si{\square\meter}   \\
Blood viscosity &   $\mu^b$     &      $4.0\times 10^{-3}$ &   \si{\pascal\second}   \\
Initial vascular porosity  &    $\varepsilon^b_0$    &  0\% or 2\% or 4\%   & -  \\
Vascular porosity  &   $\varepsilon^b $     &   Equation \ref{eq:48}    & - \\ \hline
\end{tabular}%
\caption{Mechanical parameters for the bi-compartment model}
\label{tab:5}
\end{table}

\subsection{Governing Equations}

Let one consider a vascular multi-compartment structure composed of a solid scaffold filled with interstitial fluid (IF) and blood. The medium is assumed saturated. The following convention is assumed: $\bullet^s$ denotes the solid phase, $\bullet^l$ denotes the interstitial fluid phase (IF) and $\bullet^b$ denotes the vascular part. The primary variables of the problem are the pressure applied in the pores of the extra-vascular part of the porous medium, namely $p^l$, the blood pressure, namely $p^b$, and the displacement of the solid scaffold, namely $\mathbf{u}^s$. (Equation \ref{eq:36}) links the different volume fractions. The volume fraction of the phase $\alpha$ is defined by (Equation \ref{eq:2}). $\varepsilon^l$ is called the extra-vascular porosity of the medium.

\begin{align}
 \varepsilon^s+\varepsilon^l+\varepsilon^b=1 \label{eq:36}
\end{align}

Assuming that there is no inter-phase mass transport (\textit{i.e.} the IF and the blood are assumed pure phases), the continuity equations (mass conservation) of the solid, the IF and the blood phases are respectively given by Equation \ref{eq:37}, \ref{eq:38}, \ref{eq:39}.

\begin{align}
 \frac{\partial}{\partial t}(\rho^s(1-\varepsilon^l-\varepsilon^b))+\nabla\cdot(\rho^s(1-\varepsilon^l-\varepsilon^b)\mathbf{v^s}) = 0 \label{eq:37}\\
 \frac{\partial}{\partial t}(\rho^l\varepsilon^l)+\nabla\cdot(\rho^l\varepsilon^l \mathbf{v}^l) = 0\label{eq:38}\\
 \frac{\partial}{\partial t}(\rho^b\varepsilon^b)+\nabla\cdot(\rho^b\varepsilon^b \mathbf{v}^b) = 0
 \label{eq:39}
\end{align}

According to section \ref{sec:2.1}, and dividing each equation by the corresponding density, the continuity equations can be re-expressed as:

\begin{align}
 \frac{\mathrm{D}^s}{\mathrm{D}t}(1-\varepsilon^l-\varepsilon^b)+(1-\varepsilon^l-\varepsilon^b)\nabla\cdot\mathbf{v^s} = 0 \label{eq:40}\\
 \frac{\mathrm{D}^s\varepsilon^l}{\mathrm{D}t}+\nabla\cdot(\varepsilon^l (\mathbf{v}^l-\mathbf{v}^s)) + \varepsilon^l \nabla\cdot\mathbf{v^s} = 0
 \label{eq:41}\\
  \frac{\mathrm{D}^s\varepsilon^b}{\mathrm{D}t}+\nabla\cdot(\varepsilon^b (\mathbf{v}^b-\mathbf{v}^s)) + \varepsilon^b \nabla\cdot\mathbf{v^s} = 0\label{eq:42}
\end{align}

For the fluid phase, Darcy's law (Equation \ref{eq:43}, \ref{eq:44}) is used to evaluate the fluid flow in the porous medium. 

\begin{align}
 \varepsilon^l(\mathbf{v}^l-\mathbf{v}^s) = - \frac{k^\varepsilon}{\mu^l}(\mathbf{\nabla}p^l-\rho^l\mathbf{g})\label{eq:43}\\
  \varepsilon^b(\mathbf{v}^b-\mathbf{v}^s) = - \frac{k^b}{\mu^b}(\mathbf{\nabla}p^b-\rho^b\mathbf{g}) \label{eq:44}
\end{align}

{\noindent Where $k^\varepsilon$, $k^b$ are the intrinsic permeabilities (\si{\square \meter}), $\mu^l$, $\mu^b$ are the dynamic viscosities (\si{\pascal \second}), $p^l$, $p^b$ the pressures and $\mathbf{g}$ the gravity.}

Equation \ref{eq:39} gives the following relationship: 

\begin{align}
 \frac{\mathrm{D}^s\varepsilon^l}{\mathrm{D}t}=-\frac{\mathrm{D}^s\varepsilon^b}{\mathrm{D}t}+(1-\varepsilon^l-\varepsilon^b)\nabla\cdot\mathbf{v^s} \label{eq:45}
\end{align}

Considering Equations \ref{eq:43}, \ref{eq:45}, Equation \ref{eq:41} becomes:

\begin{align}
 -\frac{\mathrm{D}^s\varepsilon^b}{\mathrm{D}t}-\nabla\cdot(\frac{k^\varepsilon}{\mu^l}\mathbf{\nabla}p^l) +(1-\varepsilon^b)\nabla\cdot\mathbf{v^s}= 0
 \label{eq:46}
\end{align}

Then, reading Equation \ref{eq:44}, Equation \ref{eq:42} gives:
\begin{align}
    \frac{\mathrm{D}^s\varepsilon^b}{\mathrm{D}t}- \nabla\cdot(\frac{k^b}{\mu^b}\mathbf{\nabla}p^b) + \varepsilon^b \nabla\cdot\mathbf{v^s} = 0\label{eq:47}
\end{align}

Considering a vascular tissue, we assume that the blood vessels are mostly surrounded by IF so they have weak direct interaction with the solid scaffold. Furthermore, the vessels are assumed compressible. Therefore, a state equation for the volume fraction of blood is introduced Equation \ref{eq:48}.

\begin{align}
    \varepsilon^b = \varepsilon^b_0 \cdot \left( 1 - \frac{p^l-p^b}{K^{\nu}}\right)\label{eq:48}
\end{align}

\noindent Where $\varepsilon^b_0$ denotes the blood volume fraction when $p^l=p^b$, $K^{\nu}$ is the vessel compressibility.

It follows that Equations \ref{eq:46}, \ref{eq:47} can be re-written as:

\begin{align}
 -\frac{\varepsilon^b_0}{K^{\nu}}\left(\frac{\mathrm{D}^s p^l}{\mathrm{D}t}-\frac{\mathrm{D}^s p^b}{\mathrm{D}t}\right)-\nabla\cdot(\frac{k^\varepsilon}{\mu^l}\mathbf{\nabla}p^l) +(1-\varepsilon^b)\nabla\cdot\mathbf{v^s}= 0
 \label{eq:49}\\
    \frac{\varepsilon^b_0}{K^{\nu}}\left(\frac{\mathrm{D}^s p^l}{\mathrm{D}t}-\frac{\mathrm{D}^s p^b}{\mathrm{D}t}\right) - \nabla\cdot(\frac{k^b}{\mu^b}\mathbf{\nabla}p^b) + \varepsilon^b \nabla\cdot\mathbf{v^s} = 0\label{eq:50}
\end{align}

Once the continuity equations are settled, one can define the quasi-static momentum balance of the porous medium, Equation \ref{eq:51}.

\begin{equation}
    \mathbf{\nabla}\cdot\mathbf{t}^{\text{tot}} = 0
    \label{eq:51}
\end{equation}

\noindent Where $\mathbf{t}^{\text{tot}}$ is the total Cauchy stress tensor. We introduce an effective stress tensor denoted $\mathbf{t}^\text{eff}$, responsible for all deformation of the solid scaffold. Then, $\mathbf{t}^{\text{tot}}$ can be expressed as:

\begin{align}
     \mathbf{t}^{tot} = \mathbf{t}^\text{eff} - (1-\zeta)p^l\mathbf{I_d} - \zeta p^b\mathbf{I_d} \label{eq:52}\\
     \mathbf{\epsilon}(\mathbf{u})=\frac{1}{2}(\nabla\mathbf{u}+\nabla\mathbf{u}^\text{T})\label{eq:53}\\
     \mathbf{t}^\text{eff} = 2\mu\mathbf{\epsilon}(\mathbf{u}^s)+\lambda \text{tr}(\mathbf{\epsilon}(\mathbf{u}^s))\mathbf{I_d} \label{eq:54}\\
     \zeta = \varepsilon_0^b\left(1-2\frac{p^l-p^b}{K^\nu}\right)\label{eq:55}
\end{align}

Four boundaries are defined: the first one, $\Gamma_u$ has imposed displacement (Equation \ref{eq:56}), the second one $\Gamma_s$ has imposed external forces (Equation \ref{eq:57}) and $\Omega_p$ has imposed pressure (fluid leakage condition (Equation \ref{eq:58}, \ref{eq:59})). We obtain:

\begin{align}
 \mathbf{t}^\text{eff} = \mathbf{t}^\text{imposed}~\text{on}~\Gamma_s \label{eq:56}\\
 \mathbf{u}^s=\mathbf{u}^\text{imposed}~\text{on}~\Gamma_u \label{eq:57}\\
 p^l=0~\text{on}~\Gamma_{p}\label{eq:58}\\
 p^b=0~\text{on}~\Gamma_{p}\label{eq:59}
\end{align}

The initial conditions are given Table \ref{tab:6}.

\begin{table}[ht!]
\centering
\begin{tabular}{llll}
\hline
Parameter & Symbol & Value & Unit \\ \hline
     Displacement     &   $\mathbf{u}^s$     &  0     &   \si{\meter}   \\
     Displacement at previous step     &   $\mathbf{u}^s_n$     &  0     &   \si{\meter}   \\
     IF pressure    &     $p^l$    &  0     &     \si{\pascal}  \\
     IF pressure at previous step   &     $p^l_n$    &  0     &     \si{\pascal}  \\
     Blood pressure    &     $p^b$    &  0     &     \si{\pascal}  \\
      Blood pressure  at previous time step  &     $p^b$    &  0     &     \si{\pascal}  \\
      Vascular porosity &     $\varepsilon^b$    &  $\varepsilon^b_0$     &    -  \\ \hline
\end{tabular}%
\caption{Initial conditions for the bi-compartment model}
\label{tab:6}
\end{table}

\subsection{Variational Form}

For the computation of the FE model, the variational form of Equation \ref{eq:49}-\ref{eq:51} must be introduced. Let one consider ($q^l$,$q^b$,v) the test functions defined in the mixed space $\text{L}_0^2(\Omega)\times\text{L}_0^2(\Omega)\times[\text{H}^1(\Omega)]^3$. 
With a first order approximation in time, Equation \ref{eq:49}, \ref{eq:50} gives:

\begin{align}
    \begin{split}  
    -\frac{\varepsilon^b_0}{K^{\nu}}\frac{1}{dt}\int_{\Omega} (p^b-p^b_n-p^l+p^l_n)q^l\text{d}\Omega+\frac{1-\varepsilon^b}{dt}\int_{\Omega} \mathbf{\nabla}\cdot(\mathbf{u}^s-\mathbf{u}^s_n)q^l\text{d}\Omega \\
    + \frac{k^\varepsilon}{\mu^l} \int_{\Omega} \mathbf{\nabla}p^l\mathbf{\nabla}q^l\text{d}\Omega = 0, \forall~q^l\in~\text{L}_0^2(\Omega)
    \end{split}
    \label{eq:60}\\
    \begin{split}  
    \frac{\varepsilon^b}{K^{\nu}}\frac{1}{dt}\int_{\Omega} (p^b-p^b_n-p^l+p^l_n)q^b\text{d}\Omega+\frac{\varepsilon^b}{dt}\int_{\Omega} \mathbf{\nabla}\cdot(\mathbf{u}^s-\mathbf{u}^s_n)q^b\text{d}\Omega \\
    + \frac{k^b}{\mu^b} \int_{\Omega} \mathbf{\nabla}p^b\mathbf{\nabla}q^b\text{d}\Omega = 0, \forall~q^b\in~\text{L}_0^2(\Omega)
    \end{split}
    \label{eq:61}
\end{align}

Similarly, by integrating by part Equation \ref{eq:51}, and including the Neumann boundary conditions, we get:

\begin{align}
\begin{split}
        \int_{\Omega} \mathbf{t}^\text{eff}:\mathbf{\nabla}\mathbf{v}\text{d}\Omega -\int_{\Omega} (1-\zeta)p^l\mathbf{\nabla}\cdot\mathbf{v}\text{d}\Omega\\ -\int_{\Omega} \zeta p^b\mathbf{\nabla}\cdot\mathbf{v}\text{d}\Omega \\
        - \int_{\Gamma_s} \mathbf{t}^\text{imposed} \cdot \mathbf{v} \text{d}\Gamma_s=0, \forall~v\in~[\text{H}^1(\Omega)]^3
    \label{eq:62}
\end{split}
\end{align}

\subsection{FEniCSx Implementation}

This section provides the code of a multi-compartment 3D column in confined compression. In order to evaluate the FEniCSx implementation, this case is similar to the Cast3m solution proposed in \citet{Scium2021}. 3 cases are studied: avascular tissue, vascular porosity of 2\% and vascular porosity of 4\%. The load is applied as a sine ramp during 5 seconds and then sustained during 125 seconds.

The time discretization is introduced.

\begin{lstlisting}[language=python]
t, t_ramp, t_sust = 0, 5, 125            # Start time
Tf                = t_ramp+t_sust        # End time
num_steps         = 1301                 # Number of time steps
dt                = (Tf-t)/num_steps     # Time step size
\end{lstlisting}

We then introduce the material parameters according to Table \ref{tab:6}. The three cases of vascularization and Equation \ref{eq:55} are defined.

\begin{lstlisting}[language=python]
E            = Constant(mesh, ScalarType(5000))  
nu           = Constant(mesh, ScalarType(0.2))  
kepsilon_l = Constant(mesh, ScalarType(1e-14)) 
mu_l         = Constant(mesh, ScalarType(1))
lambda_m     = Constant(mesh, ScalarType(E.value*nu.value/((1+nu.value)*(1-2*nu.value))))
mu           = Constant(mesh, ScalarType(E.value/(2*(1+nu.value))))
Knu   = Constant(mesh, ScalarType(1000))     #compressibility of the vessels
mu_b     = Constant(mesh, ScalarType(0.004)) #dynamic mu_l of the blood
case=1
if case ==0:
	epsilon_b_0=Constant(mesh, ScalarType(0.00)) #initial vascular porosity
	k_b=Constant(mesh, ScalarType(2e-16))   #intrinsic permeability of vessels
	def zeta(pl,pb):
		return Constant(mesh,ScalarType(0.))
elif case ==1:
	epsilon_b_0=Constant(mesh, ScalarType(0.02)) #initial vascular porosity
	k_b=Constant(mesh, ScalarType(2e-16))   #intrinsic permeability of vessels
	def zeta(pl,pb):
		return epsilon_b_0.value*(1-2*(pl-pb)/Knu.value)
elif case ==2:
	epsilon_b_0 = Constant(mesh, ScalarType(0.04))  #initial vascular porosity
	k_b = Constant(mesh, ScalarType(4e-16))   #intrinsic permeability of vessels
	def zeta(pl,pb):
		return epsilon_b_0.value*(1-2*(pl-pb)/Knu.value)
\end{lstlisting}

Then, the integration space, boundary and initial conditions are set up for the displacement, the IF pressure and the blood pressure. 

\begin{lstlisting}[language=python]
## Mechanical loading (Terzaghi)
pinit = 200 #[Pa]
T     = Constant(mesh,ScalarType(-pinit))
## Define Mixed Space (R2,R, R) -> (u,pl, pb)
element          = VectorElement("CG", mesh.ufl_cell(), 2)
pressure_element = FiniteElement("CG", mesh.ufl_cell(), 1)
MS               = FunctionSpace(mesh, MixedElement([element,pressure_element,pressure_element]))
# Create the solution and initial spaces
X0 = Function(MS)
Xn = Function(MS)
# Create the surfacic element
ds = Measure("ds", domain=mesh, subdomain_data=facet_tag)
# compute the normals
normal = FacetNormal(mesh)
# Define the Dirichlet conditions
bcs    = []
# uy=0
facets = facet_tag.find(1)
dofs   = locate_dofs_topological(MS.sub(0).sub(1), fdim, facets)
bcs.append(dirichletbc(ScalarType(0), dofs, MS.sub(0).sub(1)))
# ux=0
facets = facet_tag.find(2)
dofs   = locate_dofs_topological(MS.sub(0).sub(0), fdim, facets)
bcs.append(dirichletbc(ScalarType(0), dofs, MS.sub(0).sub(0)))
# ux=0
facets = facet_tag.find(4)
dofs   = locate_dofs_topological(MS.sub(0).sub(0), fdim, facets)
bcs.append(dirichletbc(ScalarType(0), dofs, MS.sub(0).sub(0)))
# uz=0
facets = facet_tag.find(5)
dofs   = locate_dofs_topological(MS.sub(0).sub(2), fdim, facets)
bcs.append(dirichletbc(ScalarType(0), dofs, MS.sub(0).sub(2)))
# uz=0
facets = facet_tag.find(6)
dofs   = locate_dofs_topological(MS.sub(0).sub(2), fdim, facets)
bcs.append(dirichletbc(ScalarType(0), dofs, MS.sub(0).sub(2)))
# leakage pl=pb=0
facets = facet_tag.find(3)
dofs   = locate_dofs_topological(MS.sub(1), fdim, facets)
bcs.append(dirichletbc(ScalarType(0), dofs, MS.sub(1)))
dofs   = locate_dofs_topological(MS.sub(2), fdim, facets)
bcs.append(dirichletbc(ScalarType(0), dofs, MS.sub(2)))
# Set Initial values
# Displacement
Un_, Un_to_MS = MS.sub(0).collapse()
FUn_ = Function(Un_)
with FUn_.vector.localForm() as initial_local:
	initial_local.set(ScalarType(0.0)) 
# Update Xn for all threads
Xn.x.array[Un_to_MS] = FUn_.x.array
Xn.x.scatter_forward()
# IF Pressure
Pn_, Pn_to_MS = MS.sub(1).collapse()
FPn_ = Function(Pn_)
with FPn_.vector.localForm() as initial_local:
	initial_local.set(ScalarType(0)) 
# Update Xn for all threads
Xn.x.array[Pn_to_MS] = FPn_.x.array
Xn.x.scatter_forward()
# Blood Pressure
Pbn_, Pbn_to_MS = MS.sub(2).collapse()
FPbn_ = Function(Pbn_)
with FPbn_.vector.localForm() as initial_local:
	initial_local.set(ScalarType(0)) 
# Update Xn for all threads
Xn.x.array[Pbn_to_MS] = FPbn_.x.array
Xn.x.scatter_forward()
\end{lstlisting}

Internal variables are  required. The vessels are compressible so we include the evolution of the vascular porosity as a function representing Equation \ref{eq:48}.

\begin{lstlisting}[language=python]
# Internal variables: vascular porosity
Poro_space       = FunctionSpace(mesh,pressure_element)
poro_b   = Function(Poro_space) # vascular porosity
# Initialize
with poro_b.vector.localForm() as initial_local:
	initial_local.set(ScalarType(epsilon_b_0.value)) 
# Update
poro_b.x.scatter_forward()
poro_b.name="poro_b"
\end{lstlisting}
A xdmf file is opened to store the results.

\begin{lstlisting}[language=python]
xdmf = XDMFFile(mesh.comm, "terzaghi.xdmf", "w")
xdmf.write_mesh(mesh)
\end{lstlisting}

The test functions as well as the variational form are introduced according to Equations \ref{eq:60}, \ref{eq:61}, \ref{eq:62}.

\begin{lstlisting}[language=python]
u, pl, pb       = split(X0)
u_n, pl_n, pb_n = split(Xn)
v, ql, qb = TestFunctions(MS)
dx = Measure("dx", metadata={"quadrature_degree": 4})
F = (1-poro_b)*(1/dt)*nabla_div(u-u_n)*ql*dx + ( kepsilon_l/(mu_l) )*dot( grad(pl),grad(ql) )*dx - (epsilon_b_0/Knu)*( (1/dt)*(pb-pb_n-pl+pl_n) )*ql*dx
F += poro_b*(1/dt)*nabla_div(u-u_n)*qb*dx + ( k_b/(mu_b) )*dot( grad(pb),grad(qb) )*dx + (epsilon_b_0/Knu)*( (1/dt)*(pb-pb_n-pl+pl_n) )*qb*dx
F += inner(grad(v),teff(u))*dx - (1-zeta(pl,pb))*pl*nabla_div(v)*dx - zeta(pl,pb)*pb*nabla_div(v)*dx - T*inner(v,normal)*ds(3)
\end{lstlisting}

Finally, the problem to be solved is defined and a Newton method is used for each time step, the vascular porosity is updated and the results are stored in the xdmf file. 
\begin{lstlisting}[language=python]
dX0     = TrialFunction(MS)
J       = derivative(F, X0, dX0)
Problem = NonlinearProblem(F, X0, bcs = bcs, J = J)
solver  = nls.petsc.NewtonSolver(mesh.comm, Problem)
# Set Newton solver options
solver.atol = 5e-10
solver.rtol = 1e-11
solver.convergence_criterion = "incremental"
t = 0
for n in range(num_steps):
	t += dt
	if t < t_ramp:
		f1 = 0.5 * (1 - np.cos(np.pi*t/t_ramp))
	else:
		f1 = 1
	T.value = -200*f1
	num_its, converged = solver.solve(X0)  
	X0.x.scatter_forward()
	# Update Value
	Xn.x.array[:] = X0.x.array
	Xn.x.scatter_forward()
	# Update porosity
    poro_b.x.array[:] = epsilon_b_0.value*(1-(1/Knu.value)*(X0.x.array[Pn_to_MS]-X0.x.array[Pbn_to_MS]))
	poro_b.x.scatter_forward()
	# Save data
	__u, __pl, __pb = X0.split()
	__u.name = "Displacement"
	__pl.name = "Pressure IF"
	__pb.name = "Pressure blood"
	xdmf.write_function(__u,t)
	xdmf.write_function(__pl,t)
	xdmf.write_function(__pb,t)
	xdmf.write_function(poro_b,t)
xdmf.close()
\end{lstlisting}

\subsection{Results}

\begin{figure}[ht!]
    \centering
    \begin{subfigure}[t]{0.47\textwidth}
        \includegraphics[width=\textwidth]{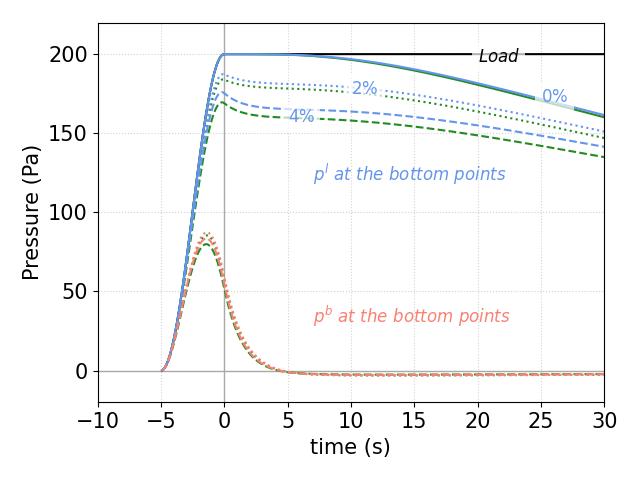}
        \caption{}
    \end{subfigure}
    \begin{subfigure}[t]{0.47\textwidth}
        \includegraphics[width=\textwidth]{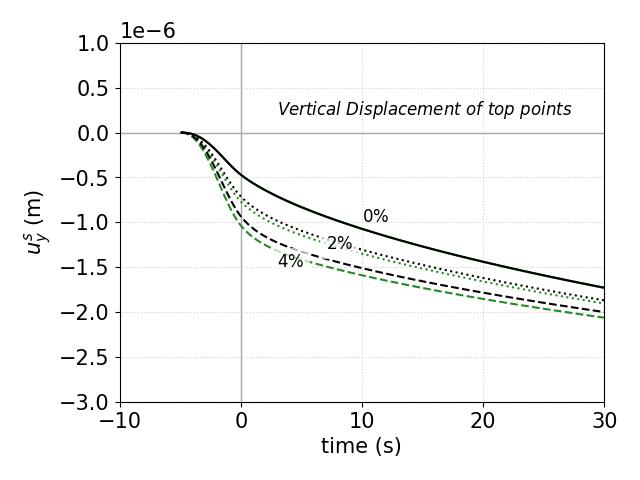}
        \caption{}
    \end{subfigure}
    \begin{subfigure}[t]{0.47\textwidth}
        \includegraphics[width=\textwidth]{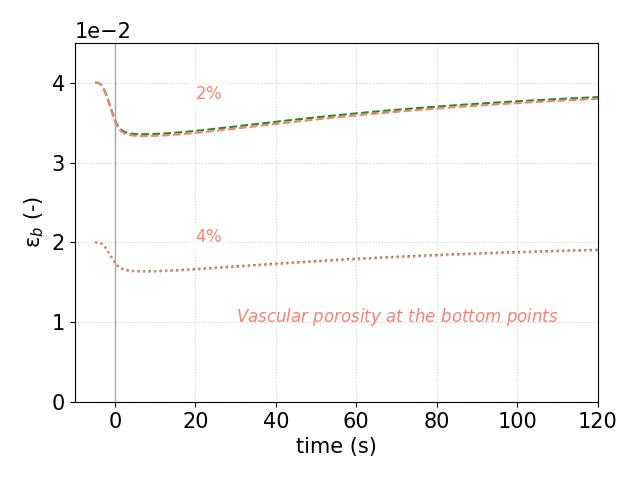}
        \caption{}
    \end{subfigure}
    \begin{subfigure}[t]{0.47\textwidth}
        \includegraphics[width=\textwidth,trim={3cm 2.5cm 3cm 2.8cm},clip]{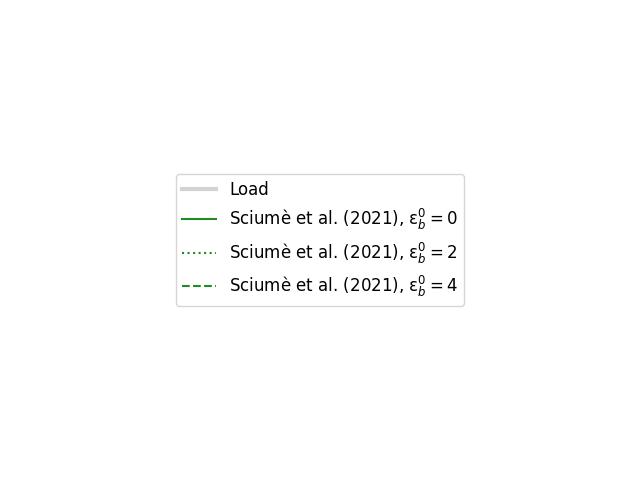}
    \end{subfigure}
    \caption{Comparison of the results obtained using FEniCSx against \citet{Scium2021} results. All results were shifted to obtain similar figures. The solid, doted and dashed lines respectively represent the 0\%, 2\% 4\% initial vascular porosity. (a) Evolution of the pressure at the bottom points. (b) Displacement of the top points. (c) Vascular porosity at the bottom points. The behavior was well retrieved for all the cases with a NRMSE lower than 10\% for all variables according to Table \ref{tab:7}.}
    \label{fig:5}
\end{figure}

The evolution of the vascular and interstitial pressures at the bottom points and the vertical displacement at the top points are provided Figure \ref{fig:5}. Each solution was obtained in $6\pm 2$ minutes on 8 threads. The overall behavior of the interstitial fluid pressure, the blood pressure and the solid displacement were retrieved. To quantitatively assess the reliability of our implemented model, The normalized root mean square error (NRMSE, Equation \ref{eq:63}) was computed for each case with the results obtained with Cast3m in \citet{Scium2021}, Table \ref{tab:7}.

\begin{equation}
    \text{NRMSE}(x,x^\text{ref})=\frac{\sqrt{\frac{1}{N}\sum_{i\in [1,N]}(x-x^\text{ref})^2}}{\text{mean}(x^\text{ref})}
    \label{eq:63}
\end{equation}

\begin{table}[ht!]
\centering
\begin{tabular}{lccc}
\hline
Parameter & 0\%               & 2\%               & 4\%               \\ \hline
$p^l$     & 1.4 \si{\percent} & 3.1 \si{\percent} & 5.1 \si{\percent} \\
$u_y$     & 0.3 \si{\percent} & 2.2 \si{\percent} & 3.7 \si{\percent} \\
$p^b$      & - & 4.7 \si{\percent}  & 8.8 \si{\percent} \\
$\varepsilon^b_0$         & - & 0.4 \si{\percent}  & 0.6 \si{\percent} \\
\end{tabular}%
\caption{NRMSE computed for each studied variable.}
\label{tab:7}
\end{table}

The NRMSE was found lower than 10\% for all unknowns. The differences are assumed to result from the method of resolution which differs between Cast3m and FEniCSx. Indeed, the Cast3m procedure relies on a staggered solver whereas our results were obtained using a monolithic solver. The order of magnitudes of the NRMSE made us however consider our solution as trustworthy.

\section{Conclusion}

The objective of this paper was to propose a step-by-step explanation of how to implement several poro-mechanical models in FEniCSx with special attention to parallel computation. Several benchmark cases for a mixed formulation were evaluated. First, a confined column was simulated under compression. Accurate results according to the L2-norm were found compared to the analytical solution. Furthermore, the code was computed 3 times faster than in the legacy FEniCS environment. Then, a possible implementation of a hyper-elastic formulation was proposed. The model was validated using \citet{SELVADURAI2016} values. Finally, a confined bi-compartment sample was simulated. The results were compared to \citet{Scium2021} data. Small differences were observed due to the choice of the solver (staggered or monolithic) but remained acceptable. The authors hope that this paper will contribute to facilitate the use of poro-elasticity in the biomechanical engineering community. This article and its supplementary material constitute a starting point to implement their own material models at a preferred level of complexity.

\section{Supplementary material}
The python codes corresponding to the workflows and the docker file of this article are made available for 2D and 3D cases on the following link: \url{https://github.com/Th0masLavigne/Dolfinx_Porous_Media.git}.

\section{Declaration of Competing Interest}
Authors have no conflicts of interest to report.


\section{Acknowledgment}
 This research was funded in whole, or in part, by the Luxembourg National Research Fund (FNR),grant reference No. 17013182. For  the purpose of open access, the author has applied a Creative Commons Attribution 4.0 International (CC BY 4.0) license to any Author Accepted Manuscript version arising from this submission. The present project is also supported by the National Research Fund, Luxembourg, under grant No. C20/MS/14782078/QuaC. 

\clearpage

\appendix

\section{3D Terzaghi example}
\label{appendix:3D:case}

Here-after is proposed a minimal working code corresponding to the 2D case included within the text.

\begin{lstlisting}[language=python]
import numpy as np
import csv
from petsc4py          import PETSc
import dolfinx
from dolfinx           import nls
from dolfinx.io        import XDMFFile
from dolfinx.mesh      import CellType, create_box, locate_entities_boundary, locate_entities, meshtags
from dolfinx.fem       import (Constant, dirichletbc, Function, FunctionSpace, locate_dofs_topological, form, assemble_scalar)
from dolfinx.fem.petsc import NonlinearProblem
from dolfinx.geometry import BoundingBoxTree, compute_collisions, compute_colliding_cells
from petsc4py.PETSc    import ScalarType
from mpi4py            import MPI
from ufl               import (FacetNormal, Identity, Measure, TestFunctions, TrialFunction, VectorElement, FiniteElement, dot, dx, inner, grad, nabla_div, div, sym, MixedElement, derivative, split)
#
def epsilon(u):
    return sym(grad(u))
#
def teff(u):
    return lambda_m * nabla_div(u) * Identity(u.geometric_dimension()) + 2*mu*epsilon(u)
#
kmax=1e3
def terzaghi_p(x):
	p0,L=pinit,Height
	cv = permeability.value/viscosity.value*(lambda_m.value+2*mu.value)
	pression=0
	for k in range(1,int(kmax)):
		pression+=p0*4/np.pi*(-1)**(k-1)/(2*k-1)*np.cos((2*k-1)*0.5*np.pi*(x[1]/L))*np.exp(-(2*k-1)**2*0.25*np.pi**2*cv*t/L**2)
	pl=pression
	return pl
#
def L2_error_p(mesh,pressure_element,__p):
	V2 = FunctionSpace(mesh, pressure_element)
	pex = Function(V2)
	pex.interpolate(terzaghi_p)
	L2_errorp, L2_normp = form(inner(__p - pex, __p - pex) * dx), form(inner(pex, pex) * dx)
	error_localp = assemble_scalar(L2_errorp)/assemble_scalar(L2_normp)
	error_L2p = np.sqrt(mesh.comm.allreduce(error_localp, op=MPI.SUM))
	return error_L2p
#
## Create the domain / mesh
Height = 1e-4 #[m]
Width  = 1e-5 #[m]
Length = 1e-5 #[m]
mesh   = create_box(MPI.COMM_WORLD, np.array([[0.0,0.0,0.0],[Length, Width, Height]]), [8, 8, 20], cell_type=CellType.tetrahedron)
#
## Define the boundaries:
# 1 = bottom, 2 = right, 3=top, 4=left, 5=back, 6=front
boundaries = [(1, lambda x: np.isclose(x[2], 0)),
              (2, lambda x: np.isclose(x[0], Length)),
              (3, lambda x: np.isclose(x[2], Height)),
              (4, lambda x: np.isclose(x[0], 0)),
              (5, lambda x: np.isclose(x[1], Width)),
              (6, lambda x: np.isclose(x[1], 0))]
#
facet_indices, facet_markers = [], []
fdim = mesh.topology.dim - 1
for (marker, locator) in boundaries:
    facets = locate_entities(mesh, fdim, locator)
    facet_indices.append(facets)
    facet_markers.append(np.full_like(facets, marker))
facet_indices = np.hstack(facet_indices).astype(np.int32)
facet_markers = np.hstack(facet_markers).astype(np.int32)
sorted_facets = np.argsort(facet_indices)
facet_tag = meshtags(mesh, fdim, facet_indices[sorted_facets], facet_markers[sorted_facets])
#
## Time parametrization
t         = 0                # Start time
Tf        = 6                # End time
num_steps = 1000             # Number of time steps
dt        = (Tf-t)/num_steps # Time step size
#
## Material parameters
E            = Constant(mesh, ScalarType(5000))  
nu           = Constant(mesh, ScalarType(0.4))
lambda_m     = Constant(mesh, ScalarType(E.value*nu.value/((1+nu.value)*(1-2*nu.value))))
mu           = Constant(mesh, ScalarType(E.value/(2*(1+nu.value))))
rhos         = Constant(mesh, ScalarType(1))
permeability = Constant(mesh, ScalarType(1.8e-15)) 
viscosity    = Constant(mesh, ScalarType(1e-2))  
rhol         = Constant(mesh, ScalarType(1))
beta         = Constant(mesh, ScalarType(1))
porosity     = Constant(mesh, ScalarType(0.2))
Kf           = Constant(mesh, ScalarType(2.2e9))
Ks           = Constant(mesh, ScalarType(1e10))
S            = (porosity/Kf)+(1-porosity)/Ks
#
## Mechanical loading 
pinit = 100 #[Pa]
T     = Constant(mesh,ScalarType(-pinit))
#
# Create the surfacic element
ds = Measure("ds", domain=mesh, subdomain_data=facet_tag)
normal = FacetNormal(mesh)
#
# Define Mixed Space (R2,R) -> (u,p)
displacement_element  = VectorElement("CG", mesh.ufl_cell(), 2)
pressure_element      = FiniteElement("CG", mesh.ufl_cell(), 1)
MS                    = FunctionSpace(mesh, MixedElement([displacement_element,pressure_element]))
#
# Define the Dirichlet condition
# 1 = bottom: uy=0, 2 = right: ux=0, 3=top: pl=0 drainage, 4=left: ux=0
bcs    = []
# uz=0
facets = facet_tag.find(1)
dofs   = locate_dofs_topological(MS.sub(0).sub(2), fdim, facets)
bcs.append(dirichletbc(ScalarType(0), dofs, MS.sub(0).sub(2)))
# ux=0
facets = facet_tag.find(2)
dofs   = locate_dofs_topological(MS.sub(0).sub(0), fdim, facets)
bcs.append(dirichletbc(ScalarType(0), dofs, MS.sub(0).sub(0)))
# ux=0
facets = facet_tag.find(4)
dofs   = locate_dofs_topological(MS.sub(0).sub(0), fdim, facets)
bcs.append(dirichletbc(ScalarType(0), dofs, MS.sub(0).sub(0)))
# uy=0
facets = facet_tag.find(5)
dofs   = locate_dofs_topological(MS.sub(0).sub(1), fdim, facets)
bcs.append(dirichletbc(ScalarType(0), dofs, MS.sub(0).sub(1)))
# uy=0
facets = facet_tag.find(6)
dofs   = locate_dofs_topological(MS.sub(0).sub(1), fdim, facets)
bcs.append(dirichletbc(ScalarType(0), dofs, MS.sub(0).sub(1)))
# drainage p=0
facets = facet_tag.find(3)
dofs   = locate_dofs_topological(MS.sub(1), fdim, facets)
bcs.append(dirichletbc(ScalarType(0), dofs, MS.sub(1)))
#
# Create the initial and solution spaces
X0 = Function(MS)
Xn = Function(MS)
#
# Initial values
# 
Un_, Un_to_MS = MS.sub(0).collapse()
FUn_ = Function(Un_)
with FUn_.vector.localForm() as initial_local:
	initial_local.set(ScalarType(0.0)) 
#
# Update Xn
Xn.x.array[Un_to_MS] = FUn_.x.array
Xn.x.scatter_forward()
#
Pn_, Pn_to_MS = MS.sub(1).collapse()
FPn_ = Function(Pn_)
with FPn_.vector.localForm() as initial_local:
	initial_local.set(ScalarType(pinit)) 
#
# Update Xn
Xn.x.array[Pn_to_MS] = FPn_.x.array
Xn.x.scatter_forward()
#
# Variational form
# Identify the unknowns from the function
u,p    =split(X0)
u_n,p_n=split(Xn)
# Set up the test functions
v,q = TestFunctions(MS)
# Equation 17
F  = (1/dt)*nabla_div(u-u_n)*q*dx + (permeability/viscosity)*dot(grad(p),grad(q))*dx  + ( S/dt )*(p-p_n)*q*dx
# Equation 18
F += inner(grad(v),teff(u))*dx - beta * p * nabla_div(v)*dx - T*inner(v,normal)*ds(3)
# Non linear problem definition
dX0     = TrialFunction(MS)
J       = derivative(F, X0, dX0)
Problem = NonlinearProblem(F, X0, bcs = bcs, J = J)
# set up the non-linear solver
solver  = nls.petsc.NewtonSolver(mesh.comm, Problem)
# Absolute tolerance
solver.atol = 5e-10
# relative tolerance
solver.rtol = 1e-11
solver.convergence_criterion = "incremental"
#
t = 0
L2_p = np.zeros(num_steps, dtype=PETSc.ScalarType)
for n in range(num_steps):
	t += dt
	num_its, converged = solver.solve(X0)
	X0.x.scatter_forward()
	# Update Value
	Xn.x.array[:] = X0.x.array
	Xn.x.scatter_forward()
	__u, __p = X0.split()
	# Compute L2 norm for pressure
	error_L2p     = L2_error_p(mesh,pressure_element,__p)
	L2_p[n] = error_L2p
	# Solve tracking
	if mesh.comm.rank == 0:
		print(f"Time step {n}, Number of iterations {num_its}, Load {T.value}, L2-error p {error_L2p:.2e}")    
if mesh.comm.rank == 0:
	print(f"L2 error p, min {np.min(L2_p):.2e}, mean {np.mean(L2_p):.2e}, max {np.max(L2_p):.2e}, std {np.std(L2_p):.2e}")
\end{lstlisting}

\section{Local refinement}
\label{sec:4.3.1}

A 3D geometry can be meshed using the GMSH API of python (\citet{gmsh}). This allows to represent complex geometries including circle arcs. An optimized and locally refined mesh can be therefore obtained. {This example uses the method proposed in the FEniCS project tutorial \footnote{see \url{https://docs.fenicsproject.org/dolfinx/main/python/demos/demo_gmsh.html}} provided by J. Dokken and G. Wells.} 
An alternative procedure in the FEniCSx environment with local refinement is then proposed in \ref{appendix:refine}. 

\subsection{Meshing using GMSH API}
\label{gmsh}
First, the environment is initialized and the physical variables required for the box/cylinder creation are defined.
\begin{lstlisting}[language=python]
import gmsh
import numpy as np
#
gmsh.initialize()
#
# box parameters
[Length, Width, Height] = [6e-4, 2.5e-4, 4e-5]
# cylinder parameters
xc,yc,zc,dx,dy,dz, r = 6e-4/2, 0, 0, 0, 0, 4e-5, 1.5e-4
# expected dimension of the mesh
gdim = 3
\end{lstlisting}

The geometries are created using built-in functions of GMSH; potential duplicates are removed. 

\begin{lstlisting}[language=python]
# create the geometry
box      = gmsh.model.occ.addBox(0, 0, 0, Length, Width, Height)
cylinder = gmsh.model.occ.addCylinder(xc,yc,zc,dx,dy,dz, r,tag=1000,angle=np.pi)
gmsh.model.occ.synchronize()
# Remove duplicate entities and synchronize
gmsh.model.occ.removeAllDuplicates()
gmsh.model.occ.synchronize()
\end{lstlisting}

Physical groups are defined: the volumes for the 3D meshing and the surfaces for tagging. Surface groups were identified based on the coordinates of the center of mass of each surface.

\begin{lstlisting}[language=python]
surfaces, volumes = [gmsh.model.getEntities(d) for d in [ gdim-1, gdim]]
print(volumes)
# Volumes
gmsh.model.addPhysicalGroup(volumes[0][0], [volumes[0][1]], -1)
gmsh.model.setPhysicalName(volumes[0][0], -1, 'Half_Cylinder')
gmsh.model.addPhysicalGroup(volumes[1][0], [volumes[1][1]], -1)
gmsh.model.setPhysicalName(volumes[1][0], -1, 'Box')
# 1 = loading, 2 = top minus loading, 3 = bottom, 4 = left, 5 = right, 6 = Front, 7 = back
bottom_marker, front_marker, back_marker, left_marker, right_marker, top_marker, indenter_marker = 3, 6, 7, 4, 5, 2, 1
bottom, front, back, left, right, top, indenter = [],[],[],[],[],[],[]
boundaries = gmsh.model.getBoundary(volumes, oriented=False)
for boundary in boundaries:
	center_of_mass = gmsh.model.occ.getCenterOfMass(boundary[0], boundary[1])
	if np.isclose(center_of_mass[1], Width):
		back.append(boundary[1])
	elif np.isclose(center_of_mass[1], 0):
		front.append(boundary[1])
	elif np.isclose(center_of_mass[0], 0):
		left.append(boundary[1])
	elif np.isclose(center_of_mass[0], Length):
		right.append(boundary[1])
	elif np.isclose(center_of_mass[2], 0):
		bottom.append(boundary[1])
	elif np.isclose(center_of_mass[2], Height) and center_of_mass[1]>Width/3: 
		top.append(boundary[1])
	else:
		indenter.append(boundary[1])
# mark the surfaces
gmsh.model.addPhysicalGroup(boundaries[0][0], bottom, bottom_marker)
gmsh.model.setPhysicalName(boundaries[0][0], bottom_marker, 'bottom')
gmsh.model.addPhysicalGroup(boundaries[0][0], front, front_marker)
gmsh.model.setPhysicalName(boundaries[0][0], front_marker, 'front')
gmsh.model.addPhysicalGroup(boundaries[0][0], back, back_marker)
gmsh.model.setPhysicalName(boundaries[0][0], back_marker, 'back')
gmsh.model.addPhysicalGroup(boundaries[0][0], left, left_marker)
gmsh.model.setPhysicalName(boundaries[0][0], left_marker, 'left')
gmsh.model.addPhysicalGroup(boundaries[0][0], right, right_marker)
gmsh.model.setPhysicalName(boundaries[0][0], right_marker, 'right')
gmsh.model.addPhysicalGroup(boundaries[0][0], top, top_marker)
gmsh.model.setPhysicalName(boundaries[0][0], top_marker, 'top')
gmsh.model.addPhysicalGroup(boundaries[0][0], indenter, indenter_marker)
gmsh.model.setPhysicalName(boundaries[0][0], indenter_marker, 'indenter')
gmsh.model.occ.synchronize()
# Write a geo file for verification in the GMSH GUI
gmsh.write('Geom_2reelle_8EP.geo_unrolled')
\end{lstlisting}

Then, a threshold function is defined over a distance field to mesh the circular area. This allows for creating an adaptive mesh: coarse far from the circular area, refine close to it.

\begin{lstlisting}[language=python]
indenter_interface = surfaces[0][1]
distance = gmsh.model.mesh.field.add("Distance")
gmsh.model.mesh.field.setNumbers(distance, "FacesList", [indenter_interface])
# A threshold function is defined:
resolution = r/10
threshold = gmsh.model.mesh.field.add("Threshold")
gmsh.model.mesh.field.setNumber(threshold, "IField", distance)
gmsh.model.mesh.field.setNumber(threshold, "LcMin", resolution)
gmsh.model.mesh.field.setNumber(threshold, "LcMax", 5*resolution)
gmsh.model.mesh.field.setNumber(threshold, "DistMin", 0.6*r)
gmsh.model.mesh.field.setNumber(threshold, "DistMax", r)
# If several fields are defined:
minimum = gmsh.model.mesh.field.add("Min")
gmsh.model.mesh.field.setNumbers(minimum, "FieldsList", [threshold]) # add other fields in the list if needed
gmsh.model.mesh.field.setAsBackgroundMesh(minimum)
\end{lstlisting}

Finally, the options of the mesher are defined and the mesh is created.

\begin{lstlisting}[language=python]
gmsh.model.occ.synchronize()
gmsh.option.setNumber("General.Terminal",1)
gmsh.option.setNumber("Mesh.Optimize", True)
gmsh.option.setNumber("Mesh.OptimizeNetgen", True)
gmsh.model.occ.synchronize()
# gmsh.option.setNumber("Mesh.MshFileVersion", 2.0)
gmsh.option.setNumber("Mesh.MeshSizeExtendFromBoundary", 0)
gmsh.option.setNumber("Mesh.MeshSizeFromPoints", 0)
gmsh.option.setNumber("Mesh.MeshSizeFromCurvature", 0)
#
gmsh.model.mesh.generate(gdim)
gmsh.write("Mesh.msh")
gmsh.finalize()
\end{lstlisting}

\subsection{Local refinement within FEniCSx}
\label{appendix:refine}
Using GMSH API, an exact circular interface is generated. However, a similar mesh could have been obtained within FEniCSx through the approximation of the circular interface around the indenter by local refining. Here-after is proposed a minimal code for local refinement inside the circular area. 

First, the required libraries are imported and a box mesh is created.

\begin{lstlisting}[language=python]
## Librairies
import dolfinx
import numpy as np
from dolfinx.mesh import create_box, CellType, refine, locate_entities, meshtags
from dolfinx.io   import XDMFFile
from mpi4py       import MPI
#
## Box 
# Dimensions of the sample
[Length, Width, Height] = [6e-4, 2.5e-4, 4e-5]
# Discretization
[nx,ny,nz] = [30,15,8]
mesh = create_box(MPI.COMM_WORLD,np.array([[0.0,0.0,0.0],[Length, Width, Height]]), [nx,ny,nz], cell_type=CellType.tetrahedron)
\end{lstlisting}

Then a locator is introduced to identify all the edges (fdim = 1) which are part of the region we aim to refine.

\begin{lstlisting}[language=python]
def test_on_boundary(x):
	return (np.sqrt(np.power(x[0]-3e-4,2)+np.power(x[1],2))<=1.5e-4)
#
refine_boudaries = [(11, lambda x: test_on_boundary(x))]
\end{lstlisting}

Finally, a loop is performed to compute several times the refinement (\textit{np.arange(N)}), using the existing \textit{refine()} function.

\begin{lstlisting}[language=python]
for _ in np.arange(2):
	# Refinement 
	refine_indices, refine_markers = [], []
	fdim = mesh.topology.dim-2
	for (marker, locator) in refine_boudaries:
		facets = locate_entities(mesh, fdim, locator)
		refine_indices.append(facets)
		refine_markers.append(np.full_like(facets, marker))
	refine_indices = np.hstack(refine_indices).astype(np.int32)
	refine_markers = np.hstack(refine_markers).astype(np.int32)
	# indices in meshtag must be sorted
	sorted_facets_refine = np.argsort(refine_indices)
	refine_tag = meshtags(mesh, fdim, refine_indices[sorted_facets_refine], refine_markers[sorted_facets_refine])
	mesh.topology.create_entities(fdim)
	mesh = refine(mesh, refine_indices[sorted_facets_refine])
\end{lstlisting}

The facets are tagged to apply boundary conditions and the mesh is written as a .xdmf file.

\begin{lstlisting}[language=python]
def Omega_top(x):
    return np.logical_and((x[2] == Height), (np.sqrt(np.power(x[0]-3e-4,2)+np.power(x[1],2))<=1.5e-4))
#
def Omega_loading(x):
    return np.logical_and((x[2] == Height), (np.sqrt(np.power(x[0]-3e-4,2)+np.power(x[1],2))>=1.2e-4))
#
# Create the facet tags (identify the boundaries)
# 1 = loading, 2 = top minus loading, 3 = bottom, 4 = left, 5 = right, 6 = Front, 7 = back
boundaries = [(1, lambda x: Omega_loading(x)),
              (2, lambda x: Omega_top(x)),
              (3, lambda x: np.isclose(x[2], 0.0)),
              (4, lambda x: np.isclose(x[0], 0.0)),
              (5, lambda x: np.isclose(x[0], Length)),
              (6, lambda x: np.isclose(x[1], 0.0)),
              (7, lambda x: np.isclose(x[1], Width))]
# Mark them
facet_indices, facet_markers = [], []
fdim = mesh.topology.dim - 1
for (marker, locator) in boundaries:
	facets = locate_entities(mesh, fdim, locator)
	facet_indices.append(facets)
	facet_markers.append(np.full_like(facets, marker))
facet_indices = np.hstack(facet_indices).astype(np.int32)
facet_markers = np.hstack(facet_markers).astype(np.int32)
sorted_facets = np.argsort(facet_indices)
facet_tag = meshtags(mesh, fdim, facet_indices[sorted_facets], facet_markers[sorted_facets])
facet_tag.name = "facets"
# Write XDMF
mesh.topology.create_connectivity(mesh.topology.dim-1, mesh.topology.dim)
with XDMFFile(mesh.comm, "facet_tags.xdmf", "w") as xdmftag:
    xdmftag.write_mesh(mesh)
    xdmftag.write_meshtags(facet_tag)
xdmftag.close()
\end{lstlisting}

\begin{figure}
    \centering
    \begin{subfigure}[t]{0.47\textwidth}
        \includegraphics[width=\textwidth]{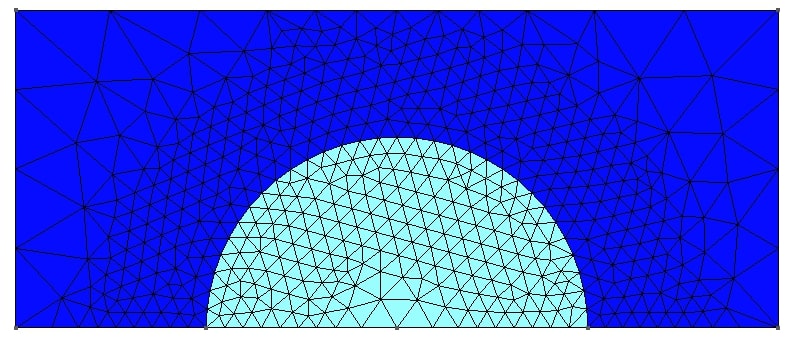}
        \caption{}
    \end{subfigure}
    \begin{subfigure}[t]{0.47\textwidth}
        \includegraphics[width=\textwidth]{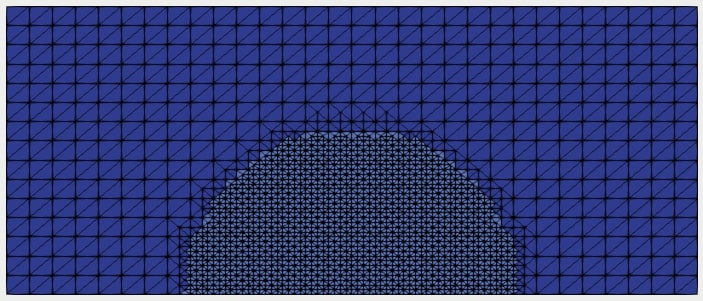}
        \caption{}
    \end{subfigure}
    \caption{GMSH (a) and FEniCSx (b) generated meshes.}
    \label{fig:B6}
\end{figure}

Figure \ref{fig:B6} gives the comparison of the mesh obtained using GMSH and the one using local refinement.

\subsection{Import an external mesh (XDMF or MSH)}
Once the mesh is generated as a tagged .msh or .xdmf file, one can consider directly read them to compile the domain and read the markers using:

\begin{lstlisting}[language=python]
from dolfinx.io.gmshio import read_from_msh
from dolfinx.io        import XDMFFile
# set value to 0 if .xdmf, set it to 1 if .msh
mesher = 1
#
if mesher == 0:
	##########################
	##  Read XDMF mesh      ##
	##########################
	filename = "filename.xdmf"
	with XDMFFile(MPI.COMM_WORLD, filename, "r") as file:
		mesh = file.read_mesh()
		mesh.topology.create_connectivity(mesh.topology.dim-1, mesh.topology.dim)
		facet_tag = file.read_meshtags(mesh, "tag.name")
#
elif mesher == 1:
	##########################
	## Read gmsh  mesh      ##
	##########################
	mesh, cell_tag, facet_tag = read_from_msh("filename.msh", MPI.COMM_WORLD, 0, gdim=3)
#
else:
	print('The mesh type is wrongly defined. mesher should equal 0 for xdmf and 1 for msh files.')
	exit()
\end{lstlisting}

\section{Evaluate the function at a physical point}
\label{appendix:eval}

One strength of using FEniCSx is its ability to evaluate the solution at given points, summing the contribution of the neighbor cells of the mesh \footnote{see \url{https://jorgensd.github.io/dolfinx-tutorial/chapter2/ns_code2.html?highlight=eval}}. 
The following code allowed to compute the figures presented for the results of the sections \ref{sec:2} and ref \ref{sec:4}.
First, one need to define the points where to evaluate the solution.

\begin{lstlisting}[language=python]
import numpy as np 
num_points = 11
y_check = np.linspace(0,Height,num_points)
points_for_time = np.array([[Width/2, 0., 0.], [Width/2, Height/2, 0.]])
points_for_space = np.zeros((num_points,3))
for ii in range(num_points):
	points_for_space[ii,0]=Width/2
	points_for_space[ii,1]=y_check[ii]
points = np.concatenate((points_for_time,points_for_space))
\end{lstlisting}

The following step is to identify the cells contributing to the points. 

\begin{lstlisting}[language=python]
from dolfinx.geometry import BoundingBoxTree, compute_collisions, compute_colliding_cells
tree = BoundingBoxTree(mesh, mesh.geometry.dim)
cell_candidates = compute_collisions(tree, points)
colliding_cells = compute_colliding_cells(mesh, cell_candidates, points)
# Here is an example to select cells contributing to the first and second points.
cells_y_0 = colliding_cells.links(0)
cells_y_H_over_2 = colliding_cells.links(1)
\end{lstlisting}

Knowing the shape of the functions to evaluate, lists are created and will be updated during the resolution procedure. Regarding parallel computation, these lists are only created on the first kernel.

\begin{lstlisting}[language=python]
from mpi4py            import MPI
if MPI.COMM_WORLD.rank == 0:
	pressure_y_0 = np.zeros(num_steps, dtype=PETSc.ScalarType)
	pressure_y_Height_over_2 = np.zeros(num_steps, dtype=PETSc.ScalarType)
	pressure_space0 = np.zeros(num_points, dtype=PETSc.ScalarType)
	pressure_space1 = np.zeros(num_points, dtype=PETSc.ScalarType)
	pressure_space2 = np.zeros(num_points, dtype=PETSc.ScalarType)
\end{lstlisting}

A function is created to evaluate a function given the mesh, the function, the contributing cells to the point and the list with its index to store the evaluated value in.

\begin{lstlisting}[language=python]
def evaluate_point(mesh, function, contributing_cells, point, output_list, index):
	from mpi4py            import MPI
	function_eval = None
	if len(contributing_cells) > 0:
		function_eval = function.eval(point, contributing_cells[:1])
	function_eval = mesh.comm.gather(function_eval, root=0)
	# Choose first pressure that is found from the different processors
	if MPI.COMM_WORLD.rank == 0:
		for element in function_eval:
			if element is not None:
				output_list[index]=element[0]
				break
	pass
\end{lstlisting}

Finally, the problem is solved for each time steps. The functions are evaluated for all kernels and gathered on the first one where the first pressure found by the different processors will be uploaded in the here-above lists. 

\begin{lstlisting}[language=python]
# time steps to evaluate the pressure in space:
n0, n1, n2 = 200,400,800
# 
t = 0
L2_p = np.zeros(num_steps, dtype=PETSc.ScalarType)
for n in range(num_steps):
	t += dt
	try:
		num_its, converged = solver.solve(X0)
	except:
		if MPI.COMM_WORLD.rank == 0:
			print("*************") 
			print("Solver failed")
			print("*************") 
			pass
	X0.x.scatter_forward()
	# Update Value
	Xn.x.array[:] = X0.x.array
	Xn.x.scatter_forward()
	__u, __p = X0.split()
	# 
	# Export the results
	__u.name = "Displacement"
	__p.name = "Pressure"
	xdmf.write_function(__u,t)
	xdmf.write_function(__p,t)
	# 
	# Compute L2 norm for pressure
	error_L2p     = L2_error_p(mesh,pressure_element,__p)
	L2_p[n] = error_L2p
	# 
	# Solve tracking
	if MPI.COMM_WORLD.rank == 0:
		print(f"Time step {n}/{num_steps}, Load {T.value}, L2-error p {error_L2p:.2e}") 
	# Evaluate the functions
	# in time
	if n == n0:
		for ii in range(num_points):
			evaluate_point(mesh, __p, colliding_cells.links(ii+2), points[ii+2], pressure_space0, ii)
		t0 = t
	elif n==n1:
			evaluate_point(mesh, __p, colliding_cells.links(ii+2), points[ii+2], pressure_space1, ii)
		t1 = t
	elif n==n2:
			evaluate_point(mesh, __p, colliding_cells.links(ii+2), points[ii+2], pressure_space2, ii)
		t2 = t
# 
xdmf.close()
\end{lstlisting}

\bibliographystyle{elsarticle-num-names} 
\bibliography{reference_corr}









 
 

\end{document}